\documentclass[sigconf]{acmart}

\setcopyright{none}
\renewcommand\footnotetextcopyrightpermission[1]{}

\usepackage{fancyhdr}
\fancypagestyle{plain}{%
   \fancyhf{} %
   \fancyfoot[L]{}%
   \fancyfoot[R]{}%
}
\fancypagestyle{firstpagestyle}{%
   \fancyhf{} %
   \fancyfoot[L]{}%
   \fancyfoot[R]{}%
}

\usepackage{subfig}
\usepackage{balance}
\usepackage{hyperref}
\usepackage{xspace}
\usepackage{graphicx, multicol}
\usepackage{datetime}
\usepackage{url}
\usepackage{graphicx}
\usepackage{color}
\usepackage{enumitem}
\usepackage{balance}
\usepackage{algorithm}
\usepackage{booktabs}
\newcommand{\name}{{MDI}\xspace}
\usepackage[noend]{algpseudocode}

\begin{document}

\title{The Case for Model-Driven Interpretability of Delay-based Congestion Control Protocols}

\author{Muhammad Khan}
\affiliation{
	\institution{New York University Abu Dhabi, UAE}
}
\email{mk7406@nyu.edu}

\author{Yasir Zaki}
\affiliation{
	\institution{New York University Abu Dhabi, UAE}
}
\email{yasir.zaki@nyu.edu}

\author{Shiva Iyer}
\affiliation{
	\institution{New York University, USA}
}
\email{shiva.iyer@nyu.edu}

\author{Talal Ahamd}
\affiliation{
	\institution{Google, USA}
}
\email{ahmad@cs.nyu.edu}

\author{Thomas Pötsch}
\affiliation{
	\institution{New York University Abu Dhabi, UAE}
}
\email{thomas.poetsch@nyu.edu}

\author{Jay Chen}
\affiliation{
	\institution{ICSI Berkeley, USA}
}
\email{jay.chen@nyu.edu}

\author{Anirudh Sivaraman}
\affiliation{
	\institution{New York University, USA}
}
\email{anirudh@cs.nyu.edu}

\author{Lakshmi Subramanian}
\affiliation{
	\institution{New York University, USA}
}
\email{lakshmi@nyu.edu}

\begin{abstract}
Analyzing and interpreting the exact behavior of new delay-based congestion control protocols with complex non-linear control loops is exceptionally difficult in highly variable networks such as cellular networks.
 This paper proposes a Model-Driven Interpretability (MDI) congestion control framework, which derives a model version of a delay-based protocol by simplifying a congestion control protocol's response into a guided random walk over a two-dimensional Markov model. We demonstrate the case for the MDI framework by using MDI to analyze and interpret the behavior of two delay-based protocols over cellular channels: Verus and Copa. Our results show a successful approximation of throughput and delay characteristics of the protocols' model versions across variable network conditions. The learned model of a protocol provides key insights into an algorithm's convergence properties.
\end{abstract}

\maketitle

\section{Introduction}
Cellular channels are known to fluctuate rapidly over short periods of time~\cite{zaki2015adaptive}. 3G and LTE network measurements~\cite{Nikravesh, Huang2013, Hu2015} demonstrated that variations in the channel cause significant performance differences across carriers, access technologies, geographic regions, and time. Rapid channel fluctuations cause loss-based congestion control (CC) algorithms to overreact and under-perform~\cite{cardwell2016bbr}, resulting in buffer-bloat and high delays~\cite{gettys2011bufferbloat, Guo2016Understanding, Jiang2012}.
Several protocols such as Sprout~\cite{winstein2013stochastic},
Copa ~\cite{copa},  Verus~\cite{zaki2015adaptive} and BBR~\cite{cardwell2016bbr} have demonstrated significant performance gains against traditional TCP variants over highly variable network channels. A common recurring theme across these protocols is to use {\em delay-based signals} to measure the network congestion state. While there is a broad array of research on the dynamics of loss-based CC protocols~\cite{padhye1998modeling,cardwell2000modeling,sun2019model}, we still lack a principled framework for understanding the dynamics of delay-based protocols.

This paper proposes {\em Model-Driven Interpretability (MDI)} CC framework, aiming to enhance the ability to interpret delay-based CC protocols' behavior. Given any protocol, the MDI framework uses empirical data on the protocol's performance for training a stochastic two-dimensional discrete-time Markov model to represent the protocol's behavior. In essence, using the empirical behavior of a protocol across diverse network conditions, MDI converts a protocol into a stochastic random walk in Markovian state space. Each state transition is determined by the delay variation feedback from the network. MDI aims to:
\begin{enumerate}[noitemsep,topsep=0pt,parsep=0pt,partopsep=0pt]
    \item Closely approximates the mean/variance of the throughput and delay distributions of the original protocol.
    \item Track the original protocol's temporal behavior, i.e., how to react to variations in network conditions.
\end{enumerate}

We note that achieving these two properties for a broad array of protocols is a non-trivial task. In the MDI framework, the notion of protocol memory is implicitly captured in the definition of the state space (transition probabilities), and the stochastic random walk using delay feedback. While the state space represents a significant approximation to the original protocol, we show that MDI successfully approximates the protocol behavior in practice.

To evaluate MDI, we developed MDI versions of two different protocols: Verus~\cite{zaki2015adaptive} and Copa~\cite{copa}. Using real-world cellular traces in 3G and 4G networks and across synthetic highly variable network conditions, we show that the MDI version of a protocol closely approximates the throughput and delay distributions of the original protocol and temporally tracks the protocols' behavior.  We demonstrate two specific benefits of MDI in this paper:

\noindent \textbf{Visualizing Protocols:}  A protocol state-space representation enables visual understanding of its behavior, including 
measuring how state transitions vary across: (i) protocols under the same network condition; (ii) network conditions under the same protocol.

\noindent \textbf{Reasoning about Convergence:}
 By representing a protocol in a Markovian state space, one can derive the mixing time and the corresponding stationary distribution of the MDI version of a protocol that we show empirically to mirror the protocol's measured statistical properties closely. 


As presented in this paper, the MDI framework is a smaller part of a much larger puzzle of understanding the properties of delay-based control protocols.
This paper has primarily shown the feasibility of the MDI framework in modeling two such protocols using a Markov Model representation. One long-term motivation to use a Markovian framework is to leverage the vast body of statistics literature on Markov models and random walks to understand the stability, dynamics, and adaptivity of delay-based protocols. While we have shown initial empirical evidence for analyzing convergence properties of protocols and visualizing protocols using MDI, a detailed statistical analysis of protocols is necessary for future work. It is beyond the scope of this paper.
\section{\name Design}
The main idea of \name is to build a model that reflects the statistical properties, providing a more intuitive and predictable understanding of the protocol behavior. At an abstract level, \name assumes that CC protocols can be modeled by the relationship between the current and the next state, where each state is a tuple of the relative change in the network delay and the sending window size.

\subsection{Modeling Delay based Control}
Consider a protocol $P$ that uses delay-variations as a congestion signal. One can imagine such a protocol maintains a recent history of delay observations, which can be used to estimates the next sending window or rate. Let us consider an epoch as the unit of time for making a decision, which can be a variable or a fixed period depending on the protocol.

The challenge in a Markov model representation of a protocol $P$ is determining the appropriate state space and mapping the protocol actions to transitions within the states. The most straightforward approach is to map the absolute values directly by describing a state as ($d_i$,$w_i$) where $d_i$ and $w_i$ are the experienced delay and sending window in an epoch $i$, respectively. We use $d$ and $w$ (without the epoch subscript $i$) to abstractly represent the observed delay and window parameters for brevity. While a two-variable state space using $(d,w)$ is simple, it may not be rich/generic since it may not be sufficient to capture the variations in these parameters. Suppose one were to represent the state space using a history of delay and window measurements. In that case, the state space representation could be much more vibrant but correspondingly much harder to learn accurately. In fact, for each additional dimension in the state space, we need an order of magnitude more training data to determine the state transitions. To capture the variations of the delay and window in the state space, we also consider (1) relative change in the delay across neighboring epochs (captured by $\alpha(d)$); (2) relative change in the window across adjacent epochs (obtained by $\beta (w)$).
These four parameters provide a richer representation of the state space. However, the training data required for the 4-dimensional space is at least two orders of magnitude more than the  $(d,w)$ space. To balance between state complexity and state richness, we chose to condense these four parameters into two composite parameters as $\alpha(d) \cdot log_{10}(d)$ and $\beta(w) \cdot log_{10}(w)$.  By representing the delay and window in log space and quantizing the values (described in Section 2.2), we can better delineate variations in relative delay (or window) changes in comparison to variations in the actual delay (or window) values across different buckets in the state space.  The quantization of these values also helps maintain a condensed two-parameter representation of the four parameters: window, delay, a relative change in window size, and the relative change in delay across epochs. We note that one can choose alternate state-space representations for the MDI framework; the key requirements are to balance the number of quantized states in the state space to capture protocol dynamics across different network conditions.

\subsection{Discrete-time Markov Model States}
A discrete-time Markov model of a protocol is represented in the form of a state-transition probability matrix. The matrix describes transition probabilities from one state to another obtained by training a protocol on a large set of network configurations. We call this the \textit{training phase} of the Markov model. To capture the protocol behavior, the matrix should include as many states as the ones observed during the training. The state is defined as a tuple with value pairs of $(\hat{d_i},\hat{w_i})$. Where $\hat{d_i}$ and $\hat{w_i}$ are calculated using the current epoch's packet delays ($d_i$) and sending-window ($w_i$) and the previous epoch's delay ($d_{i-1}$) and sending-window ($w_{i-1}$):
\begin{equation}
\hat{d_i} = \left [\left (\frac{d_{i}}{d_{i-1}}\right ) -1\right ]*log_{10}(d_{i})
\label{matrix_eq1}
\end{equation}
\begin{equation}
\hat{w_i} = \left [\left (\frac{w_{i}}{w_{i-1}}\right ) -1\right ]*log_{10}(w_{i})
\label{matrix_eq2}
\end{equation}

Assume that a protocol $P$ adjusts the congestion window as a function of delay feedback. A user executing protocol $P$ has currently the following values: the current sending window $w_i$, and the previous epoch delay feedback $d_{i-1}$. To decide on the value of the next window $w_{i+1}$, the user has to first identify the current delay $d_i$. The protocol $P$ decides the next window $w_{i+1}$ based on the following factors: the prior window $w_{i}$, and the delay variations. Only upon observing $d_i$, $P$ would be aware of the true represented state $(\hat{d_i},\hat{w_i})$ in the model space of the protocol. Essentially, given an initial window $w_i$ and delay $d_{i-1}$, the protocol $P$ has three variations that influence a transition from $(\hat{d_i},\hat{w_i})$ to $(\hat{d_{i+1}},\hat{w_{i+1}})$: (i) the variation in the initial observation $d_i$; (ii) the variation in the decision making of $w_{i+1}$; (iii) the variation in the next delay observation $d_{i+1}$.
Note that, it is not necessary for two users running the same protocol $P$ and in the same state $(\hat{d_i},\hat{w_i})$, to derive the same next window $w_{i+1}$. This decision is influenced by two factors: (i) different windows/delays values could effectively arrive at the same model state $(\hat{d_i},\hat{w_i})$; (ii) different flows may observe variations in prior observations of delays and windows. 
\subsection{Deriving the \name Transition Matrix}
The key assumption that \name makes is that the state transition from $(\hat{d_i},\hat{w_i})$ to $(\hat{d_{i+1}},\hat{w_{i+1}})$ can be captured by a guided Markov model with two basic properties: the delay feedback guides the direction of the window change (increase or decrease), and the delay variations of $d_i$ and $d_{i+1}$ have an inherent randomness that influence the protocol choice of the next window $w_{i+1}$. The guided Markov assumption is clearly an approximation of the original protocol behavior.
To derive the transition matrix, we use a protocol emulation strategy in a constrained network environment. Consider a network simulation environment where one can execute the protocol $P$ under various network conditions and background traffic. Our setup's network environment is defined by a set of network traces that specify bandwidth, packet loss, and RTT variations. 
The protocol $ P $ can be executed by simulating network flows executing the protocol in the presence of competing traffic.
We perform a broad array of network simulations by varying the network traces and the background traffic emulating several real-world protocols, including $P$. For each simulation, we measure the state transitions of $P$ across the model states. By observing all possible state transitions of $(\hat{d_i},\hat{w_i})$, with $\hat{d_i}$ ranging from $\hat{d}_{min}$ to $\hat{d}_{max}$, and $\hat{w_i}$ ranging from $\hat{w}_{min}$ to $\hat{w}_{max}$, a 2D Markov chain is created defining the following states: current state $(\hat{d_k},\hat{w_l})$ and next state $(\hat{d_r},\hat{w_v})$, where $k$ and $l$ are the current state indexes of $\hat{d_i}$ and $\hat{w_i}$, respectively. Similarly, $r$ and $v$ represents the next state indexes. To reduce the state space of possible values for $(\hat{d_i},\hat{w_i})$, we quantize these values into small buckets. \name captures the state transitions in the form of a transition probability matrix written as: 
\begin{equation}
p_{(k,l),(r,v)}=p[(\hat{d_k},\hat{w_l})|(\hat{d_r},\hat{w_v})].
\label{eq:trans_probability}
\end{equation}
Thus, $(\hat{d_k},\hat{w_l})$ defines a specific row in the transition matrix. Depending on $\hat{d_{i+1}}$ next value, represented by $r$, we obtain a subset of values from this specific row (i.e., the probability going to any of the possible $\hat{w_{i+1}}$ in the state $(\hat{d_r},\hat{w_v})$.
\subsection{Model Training Methodology}\label{sec:training}

\begin{figure}[hbt]
\centering
  \includegraphics[scale=0.22]{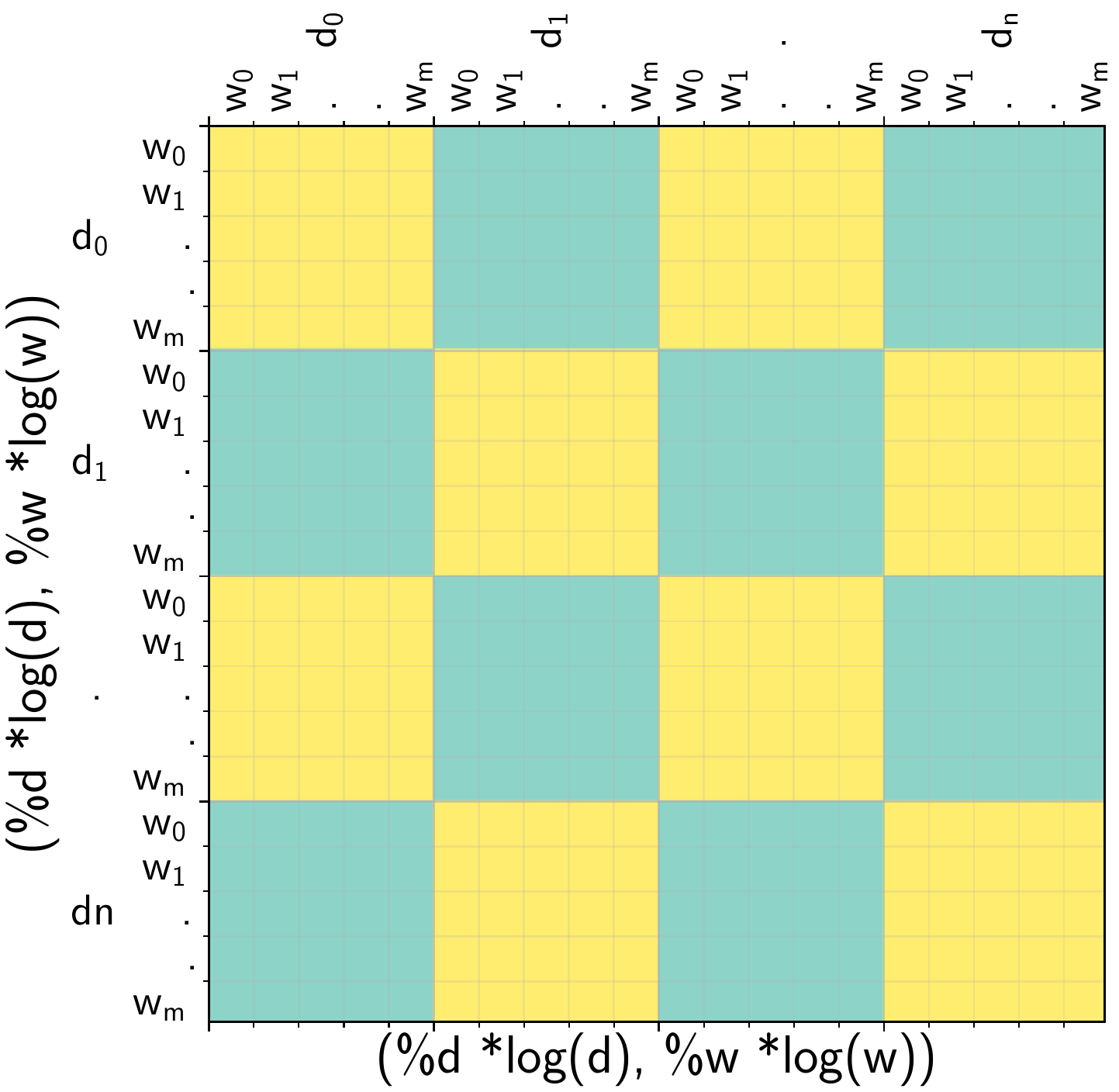}
  \caption{Transition Probability representation of a Model}
  \label{matrix}
\end{figure}

This paper focuses on training two delay-based protocols: Verus and Copa. The training is performed over a large sample of cellular traces covering a wide range of diverse scenarios. We ran each protocol through a network emulator over a large set of traces randomly synthesized from the training traces. The protocol behavior is captured by logging the set of congestion windows and their experienced correlated delays in each run. Next, the logged window and delay values are quantized (Equation~\ref{matrix_eq1} and ~\ref{matrix_eq2}). The quantized values are used to obtain the transition probability matrix where each state is the quantized pair $(\hat{w_i},\hat{d_i})$. The matrix is structured in quadrants, highlighted by yellow and green in Figure~\ref{matrix}.

Each quadrant represents a particular current delay $\hat{d_{i}}$ on the y-axis and a next delay $\hat{{d}_{i+1}}$ on the x-axis, these values are quantized in the range $d_0$ to $d_n$ to keep the matrix from being prohibitively long. Each quadrant is further divided into smaller chunks representing the current values of $\hat{w_i}$ on the y-axis and a next window variable $\hat{w_{i+1}}$ on the x-axis, which are quantized in the range $w_0$ to $w_n$. Figure~\ref{matrix} shows an empty sample matrix. 
The transition probability for each chunk is computed by counting the number of occurrences of going from one state to another as $[(\hat{d_i},\hat{w_i}),(\hat{d}_{i+1},\hat{w}_{i+1})]$. We normalize each row within a quadrant so that all outgoing transition probabilities of any state would sum to 1. 

\begin{figure*}[ht!]
    \begin{multicols}{2}
        \centering
        \begin{multicols}{3}
        \subfloat[\label{fig2:a}instantaneous]{\includegraphics[width=0.17\textwidth]{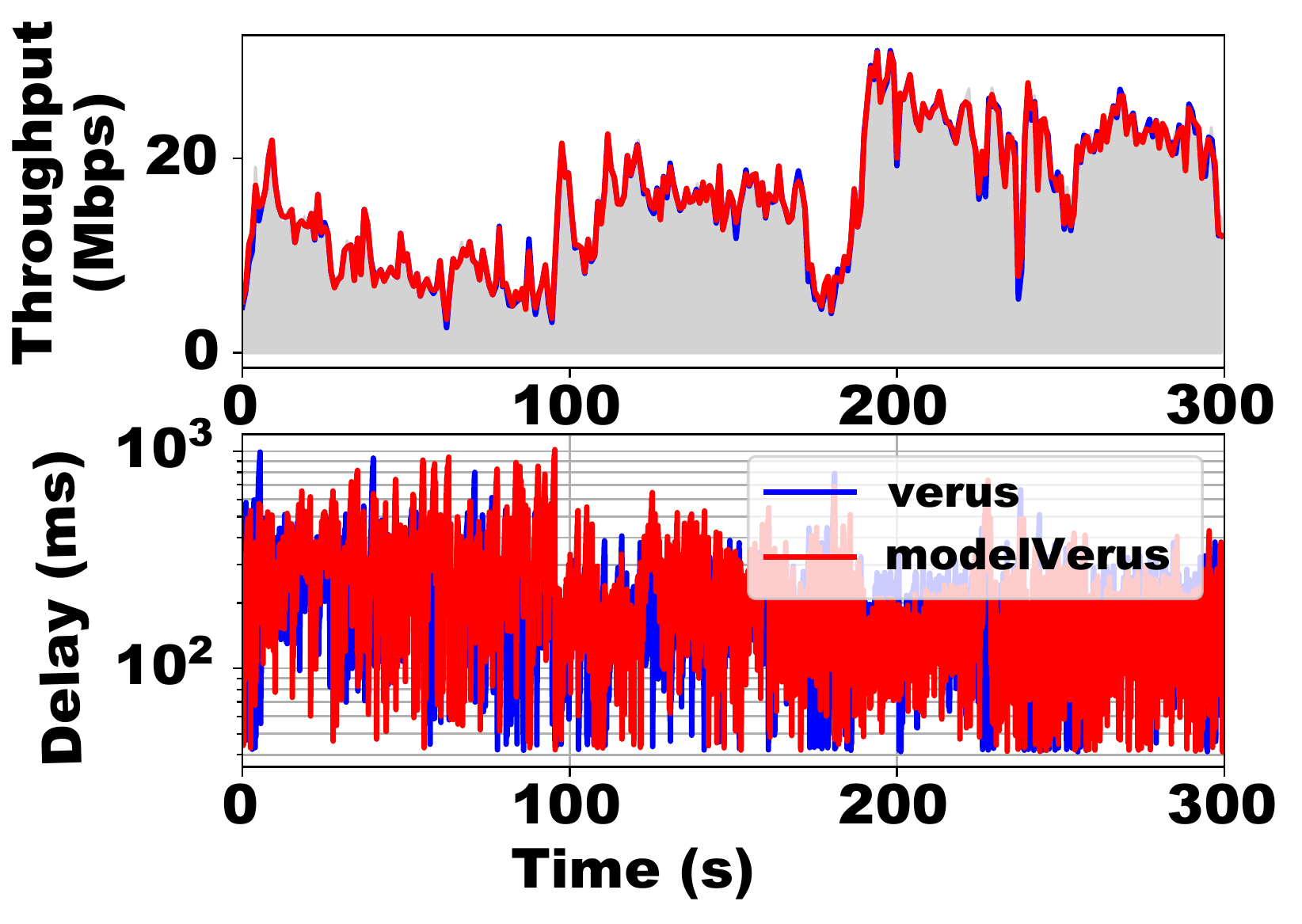}}
        \subfloat[\label{fig2:b}PDF]{\includegraphics[width=0.175\textwidth]{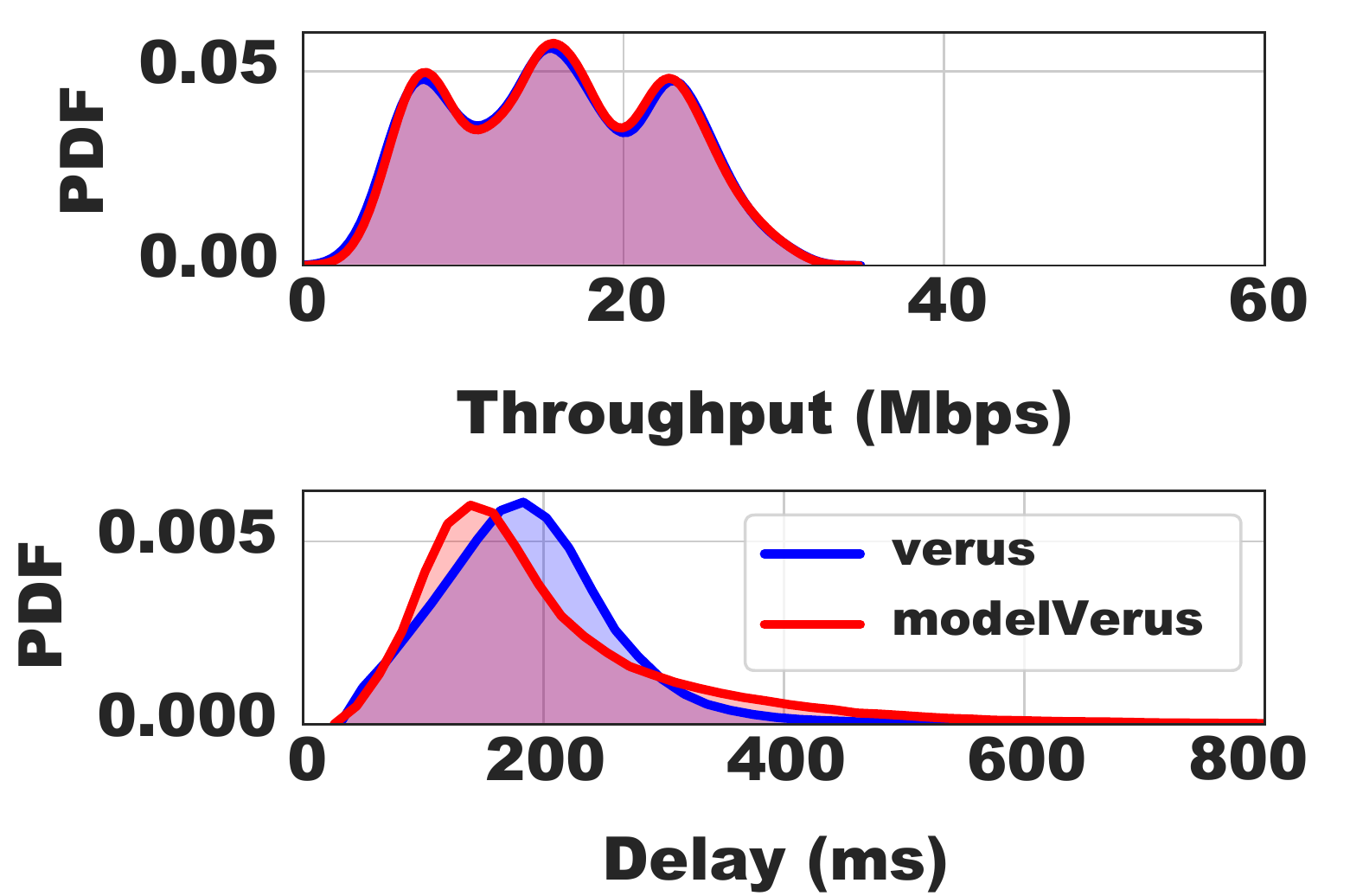}}
        \subfloat[\label{fig2:c}Population]{\includegraphics[width=0.155\textwidth]{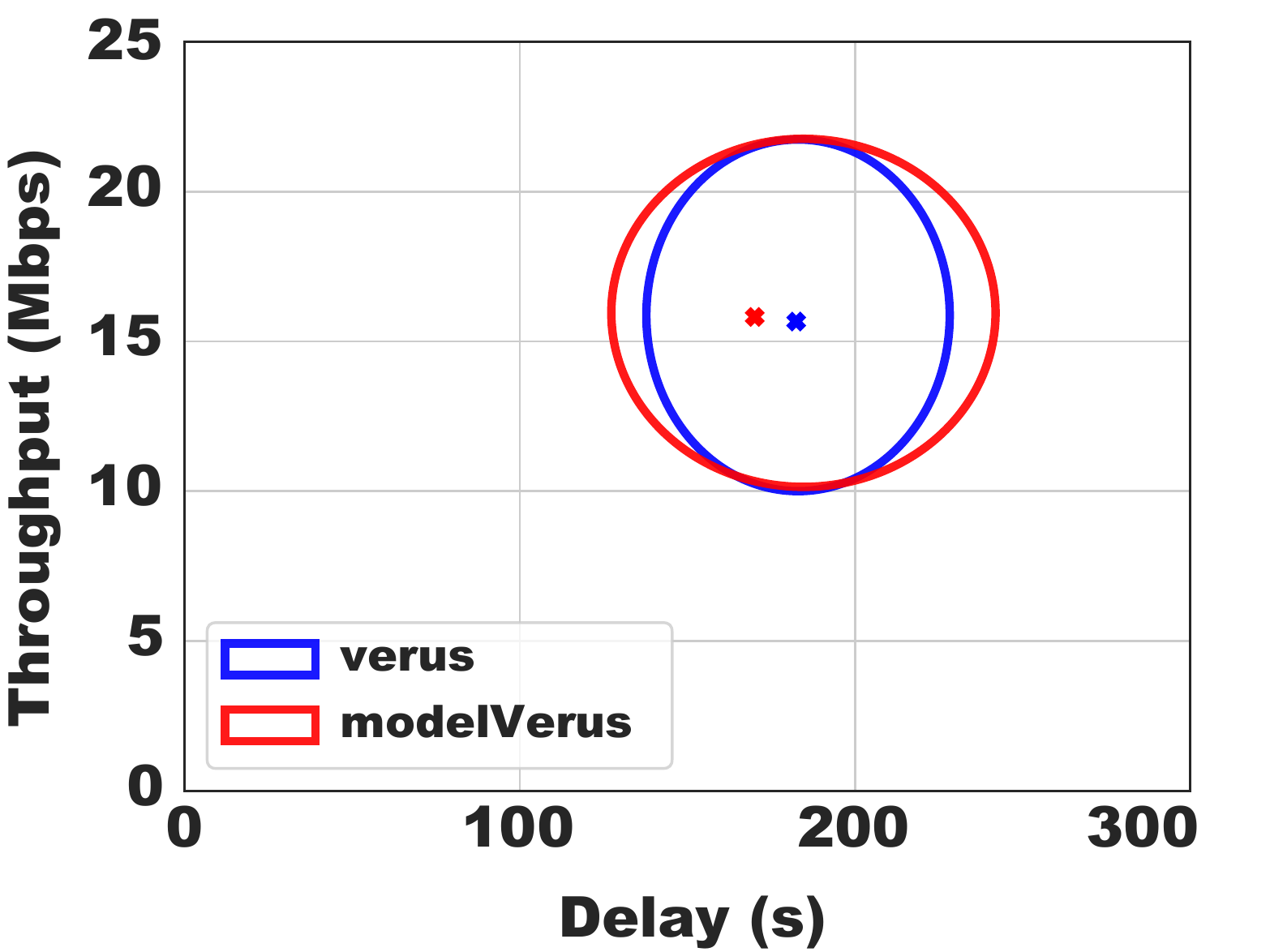}}
        \end{multicols}
        \caption{Verus Highway}
        \begin{multicols}{3}
        \subfloat[\label{fig3:a}instantaneous]{\includegraphics[width=0.17\textwidth]{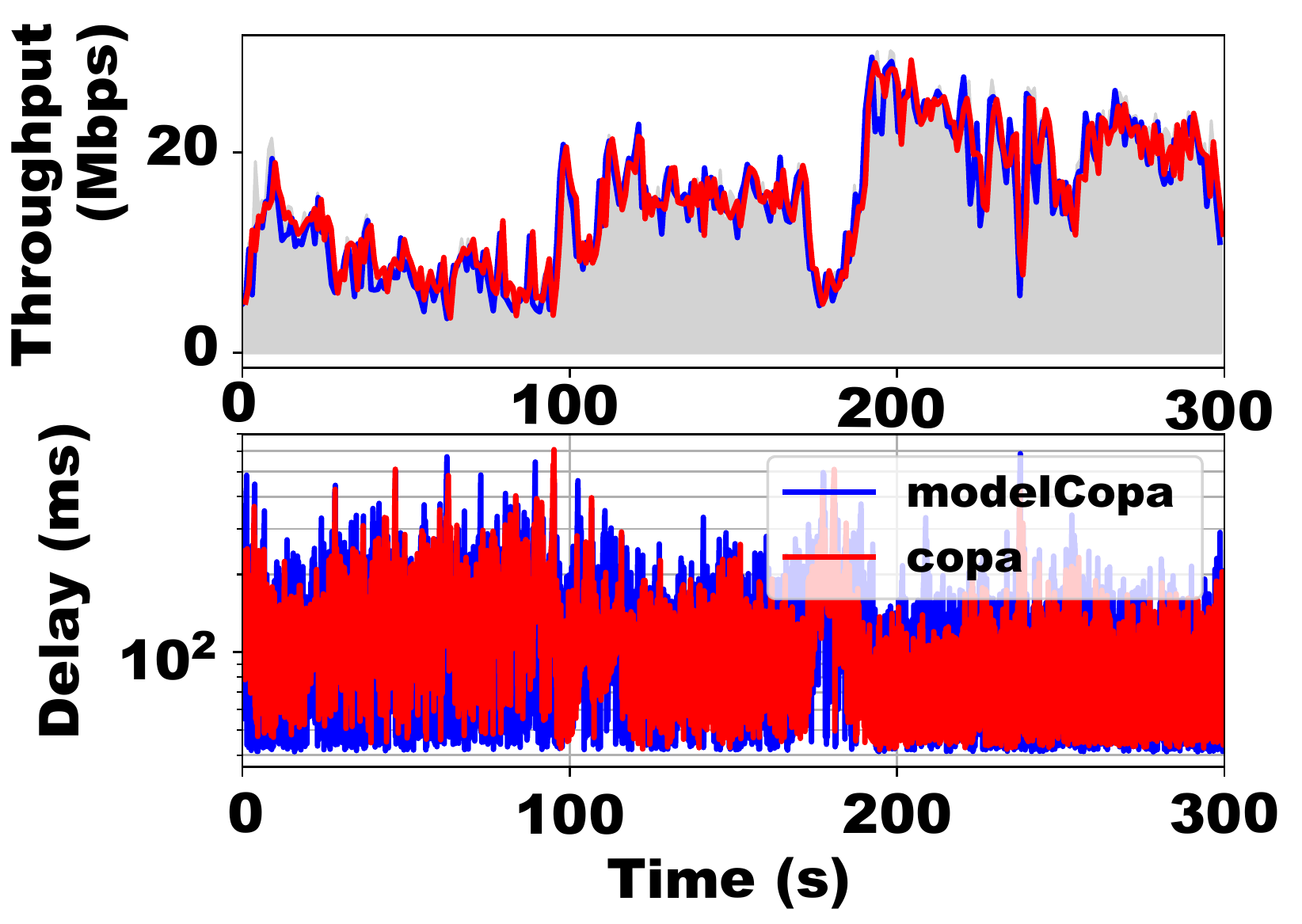}}
        \subfloat[\label{fig3:b}PDF]{\includegraphics[width=0.175\textwidth]{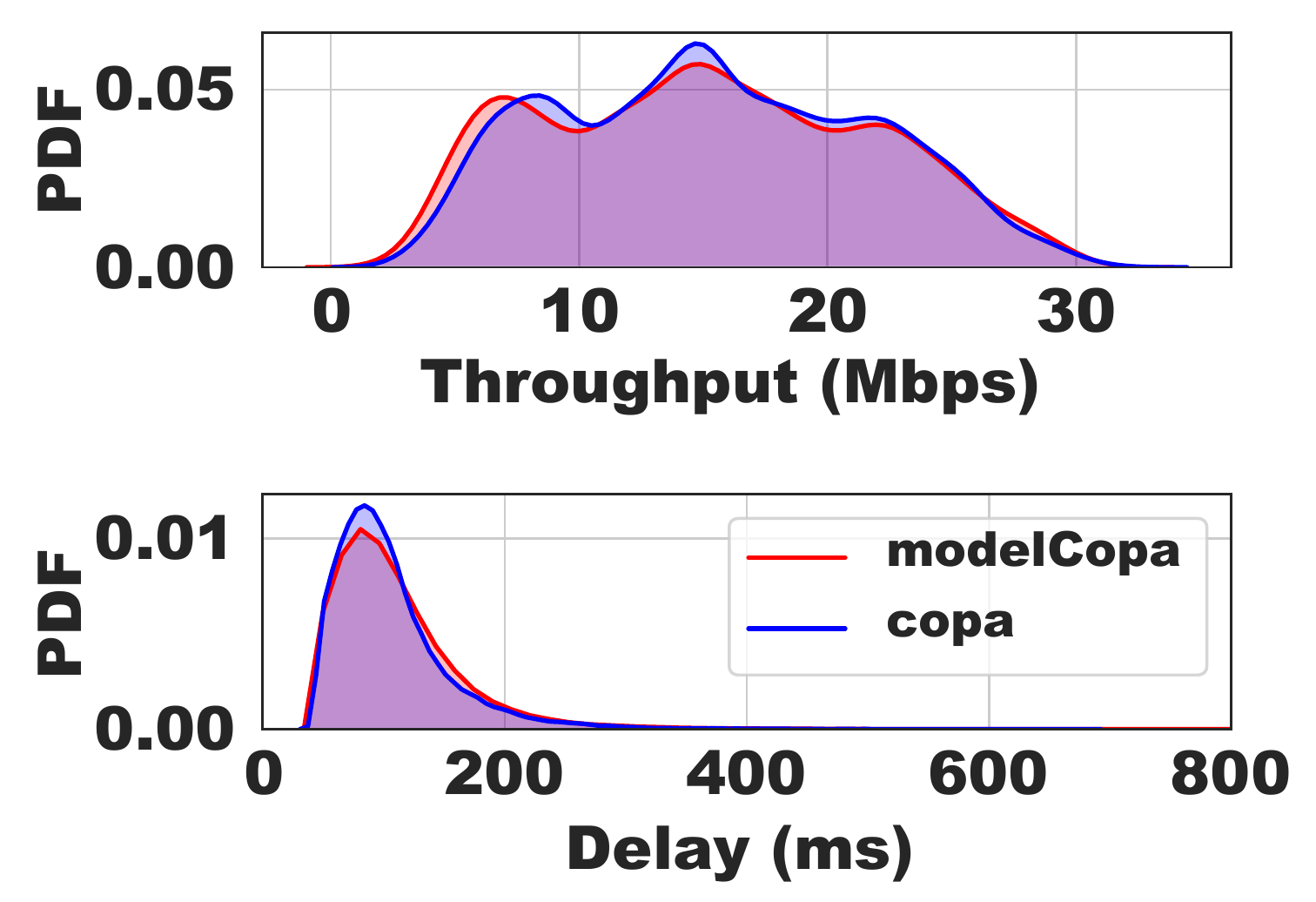}}
        \subfloat[\label{fig3:c}Population]{\includegraphics[width=0.155\textwidth]{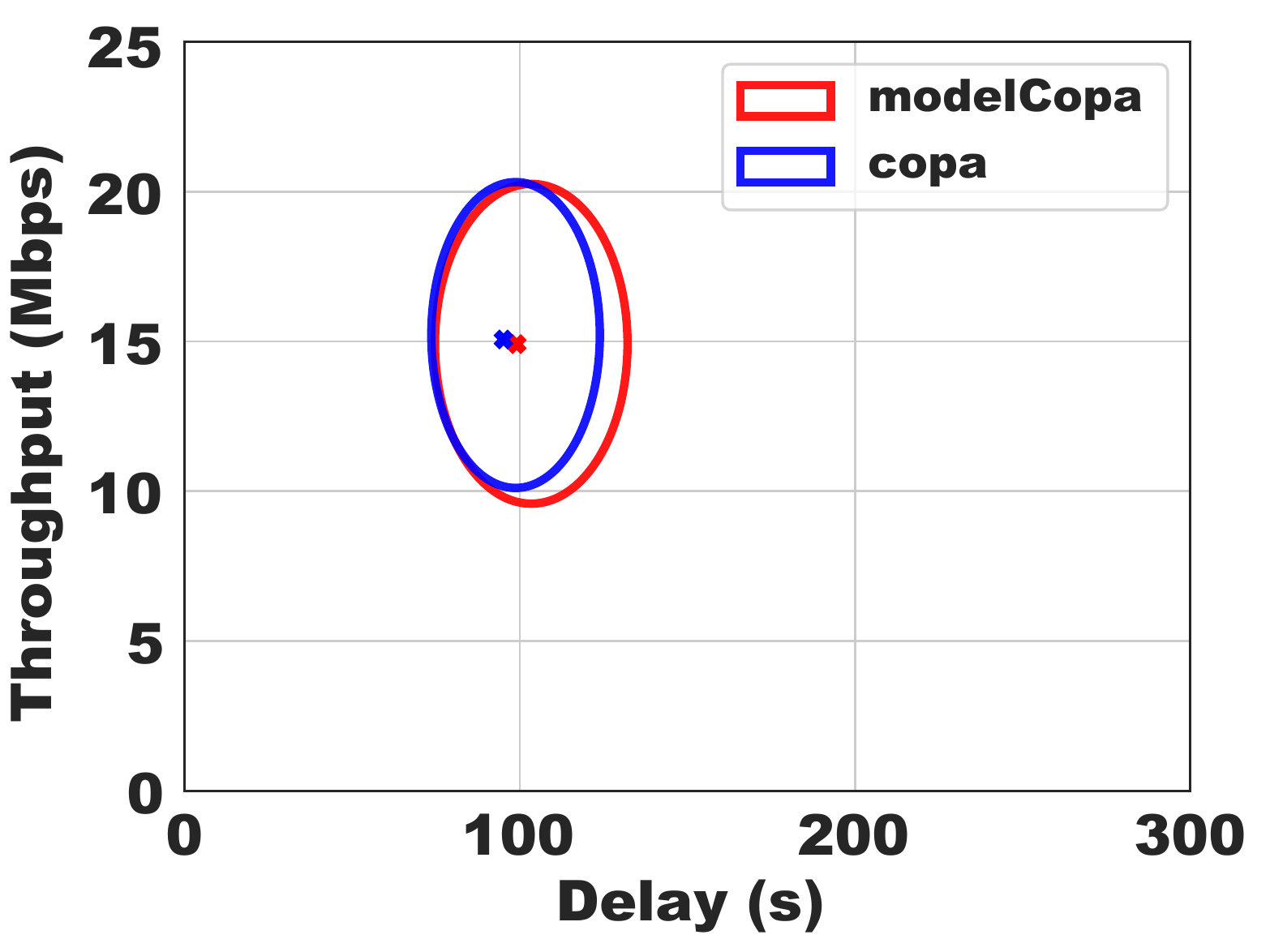}}
        \end{multicols}
        \caption{Copa Highway}
    \end{multicols}
    \begin{multicols}{2}
        \centering
        \begin{multicols}{3}
        \subfloat[\label{fig4:a}instantaneous]{\includegraphics[width=0.17\textwidth]{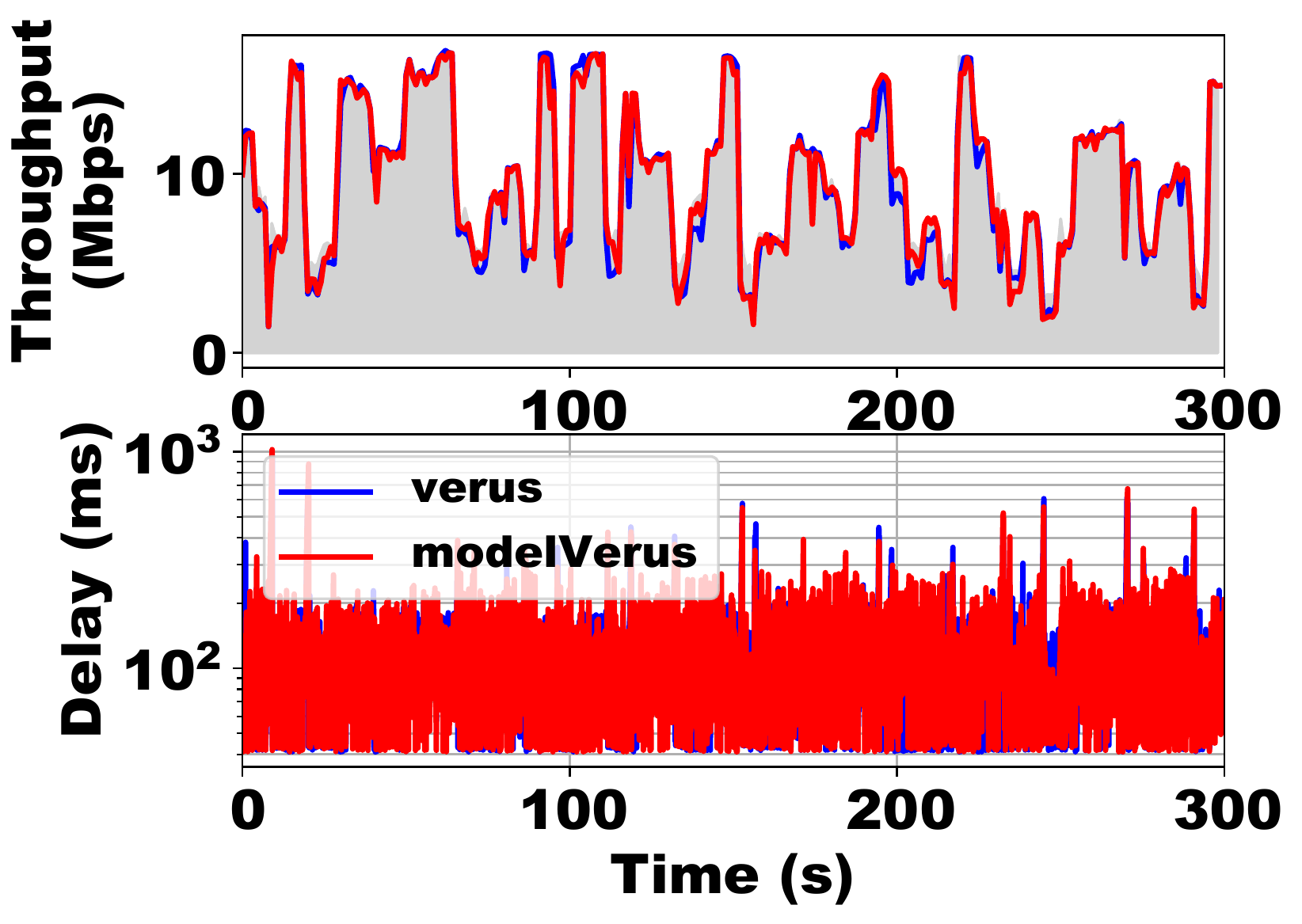}}
        \subfloat[\label{fig4:b}PDF]{\includegraphics[width=0.175\textwidth]{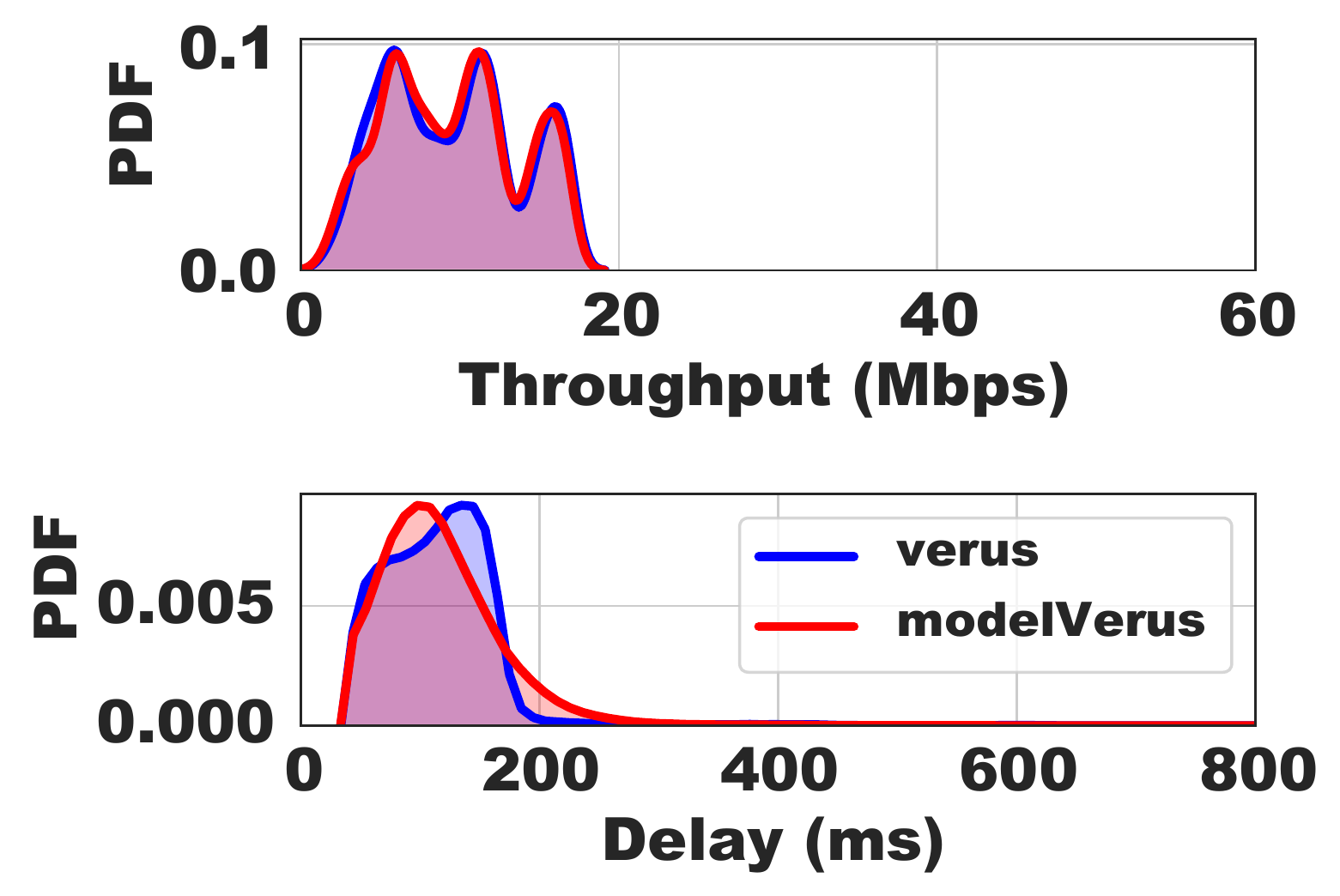}}
        \subfloat[\label{fig4:c}Population]{\includegraphics[width=0.155\textwidth]{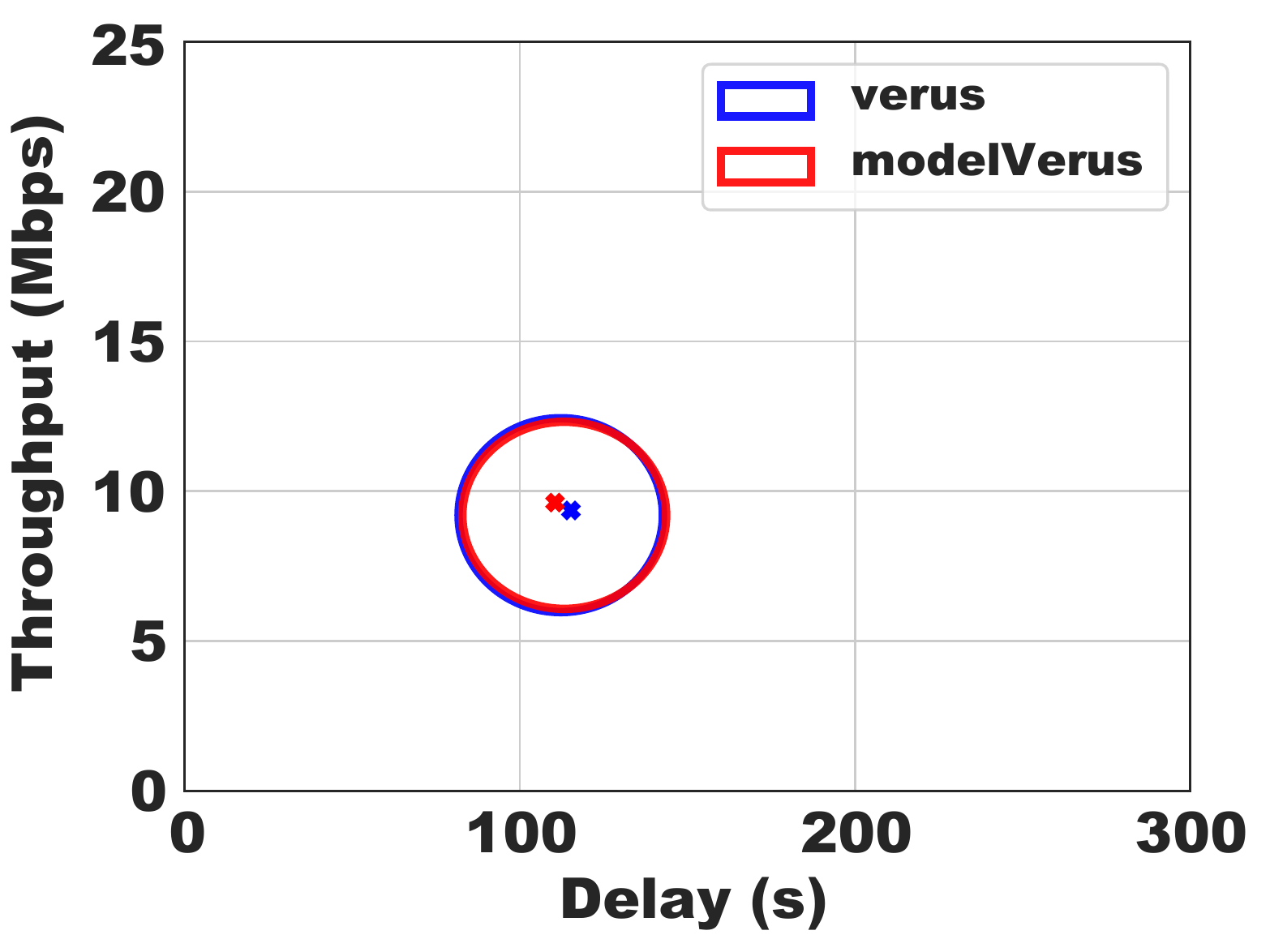}}
        \end{multicols}
        \caption{Verus Rapidly changing network}
        \begin{multicols}{3}
        \subfloat[\label{fig5:a}instantaneous]{\includegraphics[width=0.17\textwidth]{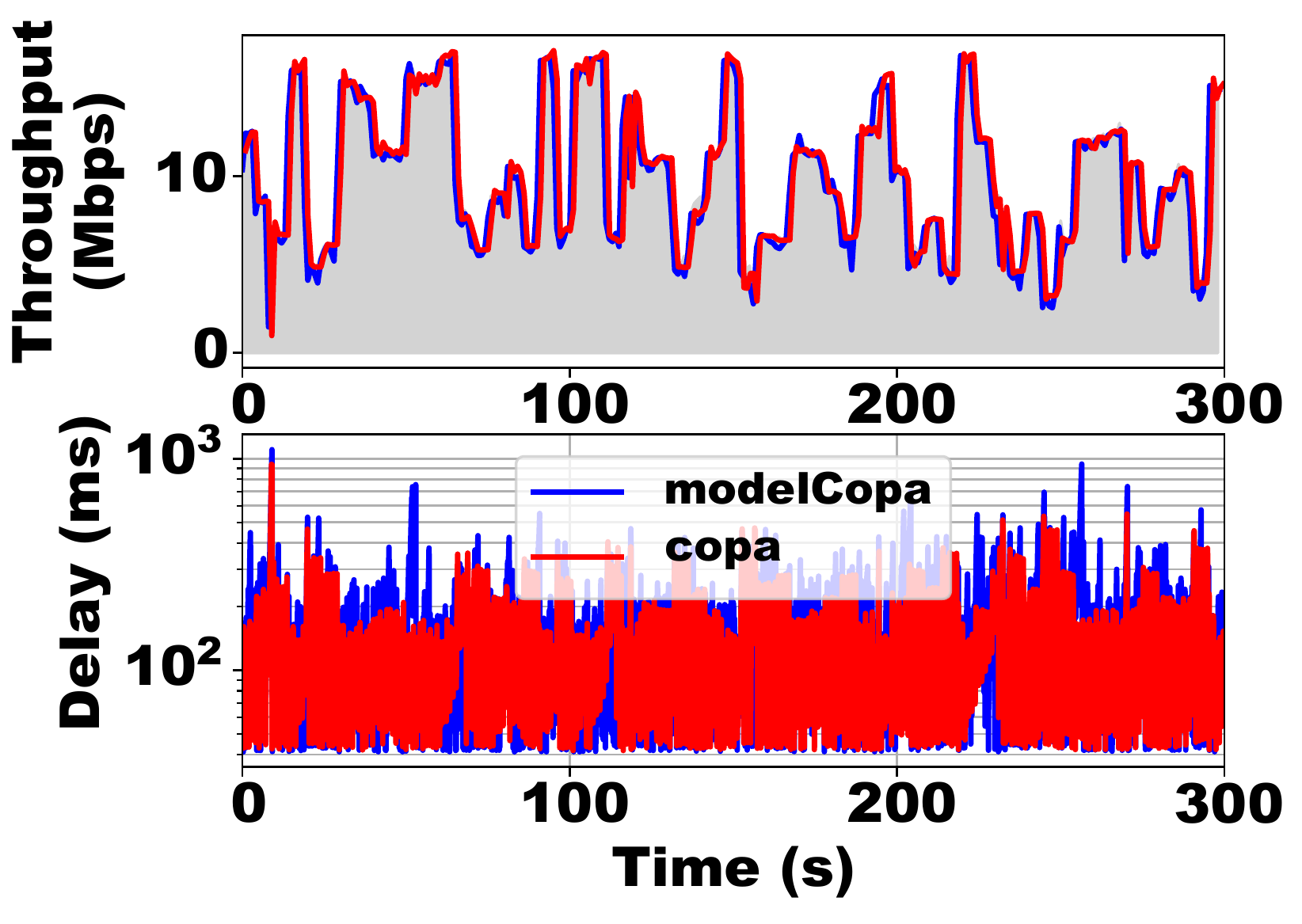}}
        \subfloat[\label{fig5:b}PDF]{\includegraphics[width=0.175\textwidth]{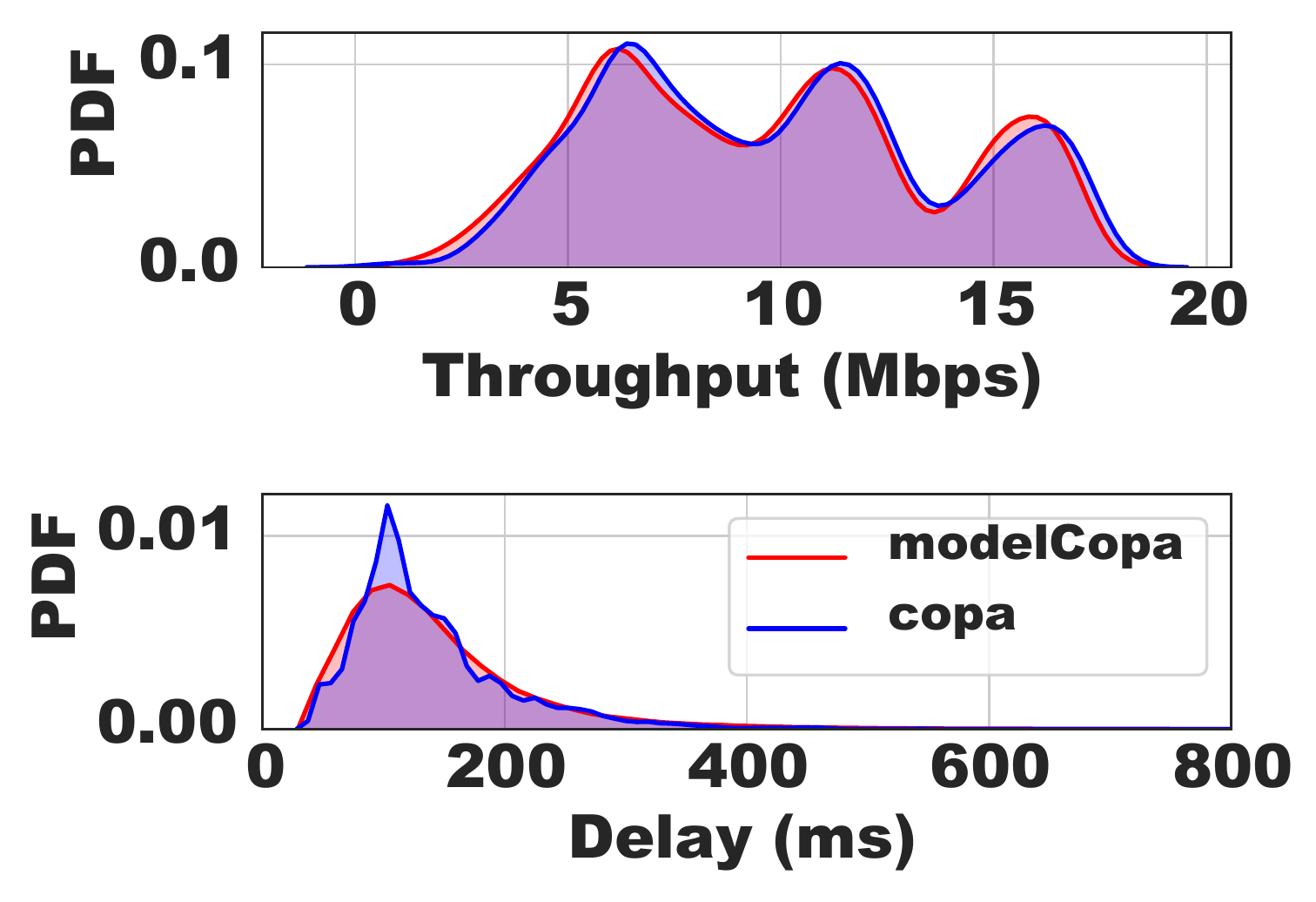}}
        \subfloat[\label{fig5:c}Population]{\includegraphics[width=0.155\textwidth]{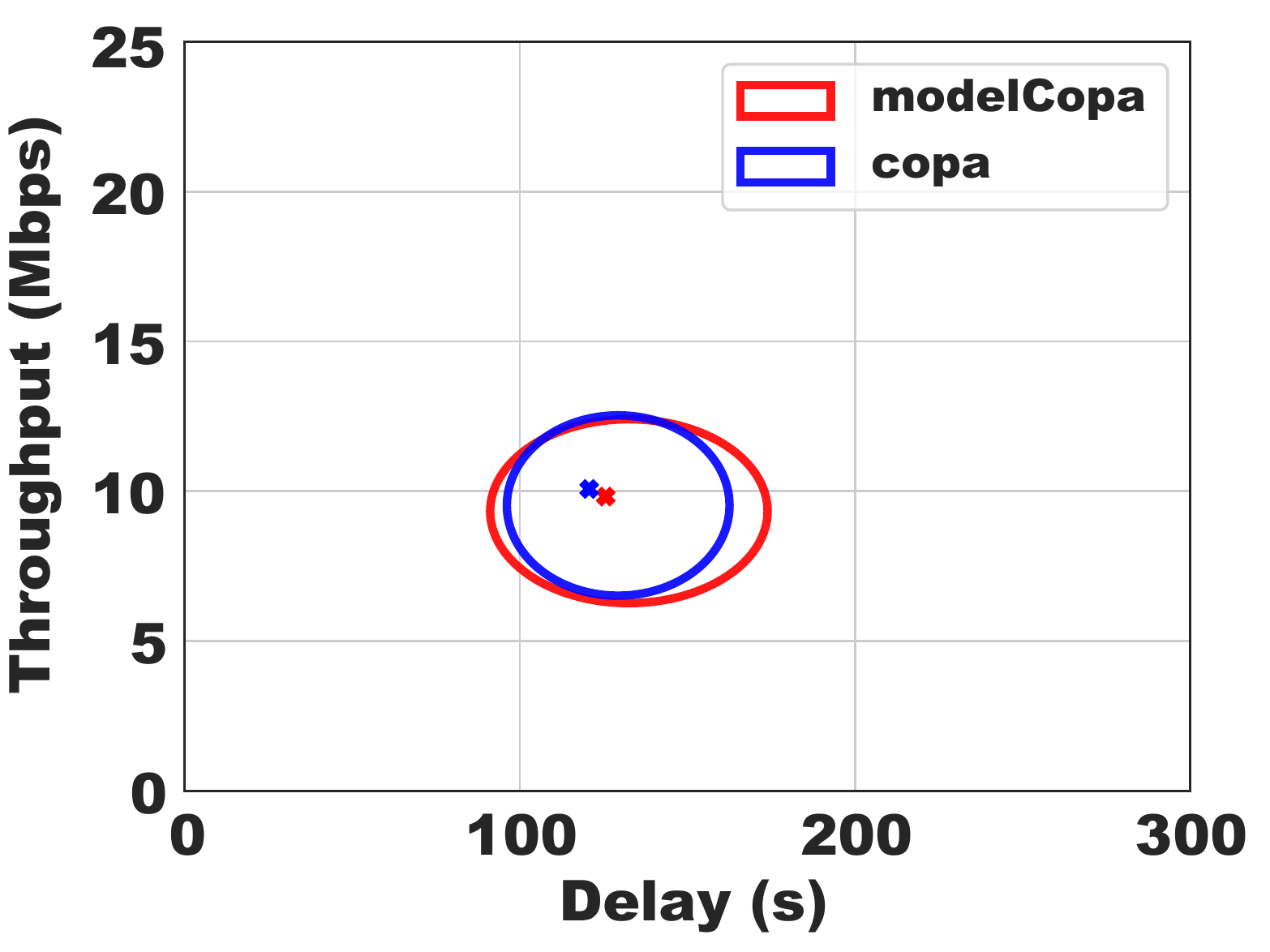}}
        \end{multicols}
        \caption{Copa Rapidly changing network}
    \end{multicols}
    \begin{multicols}{2}
        \centering
        \begin{multicols}{3}
        \subfloat[\label{fig6:a}instantaneous]{\includegraphics[width=0.17\textwidth]{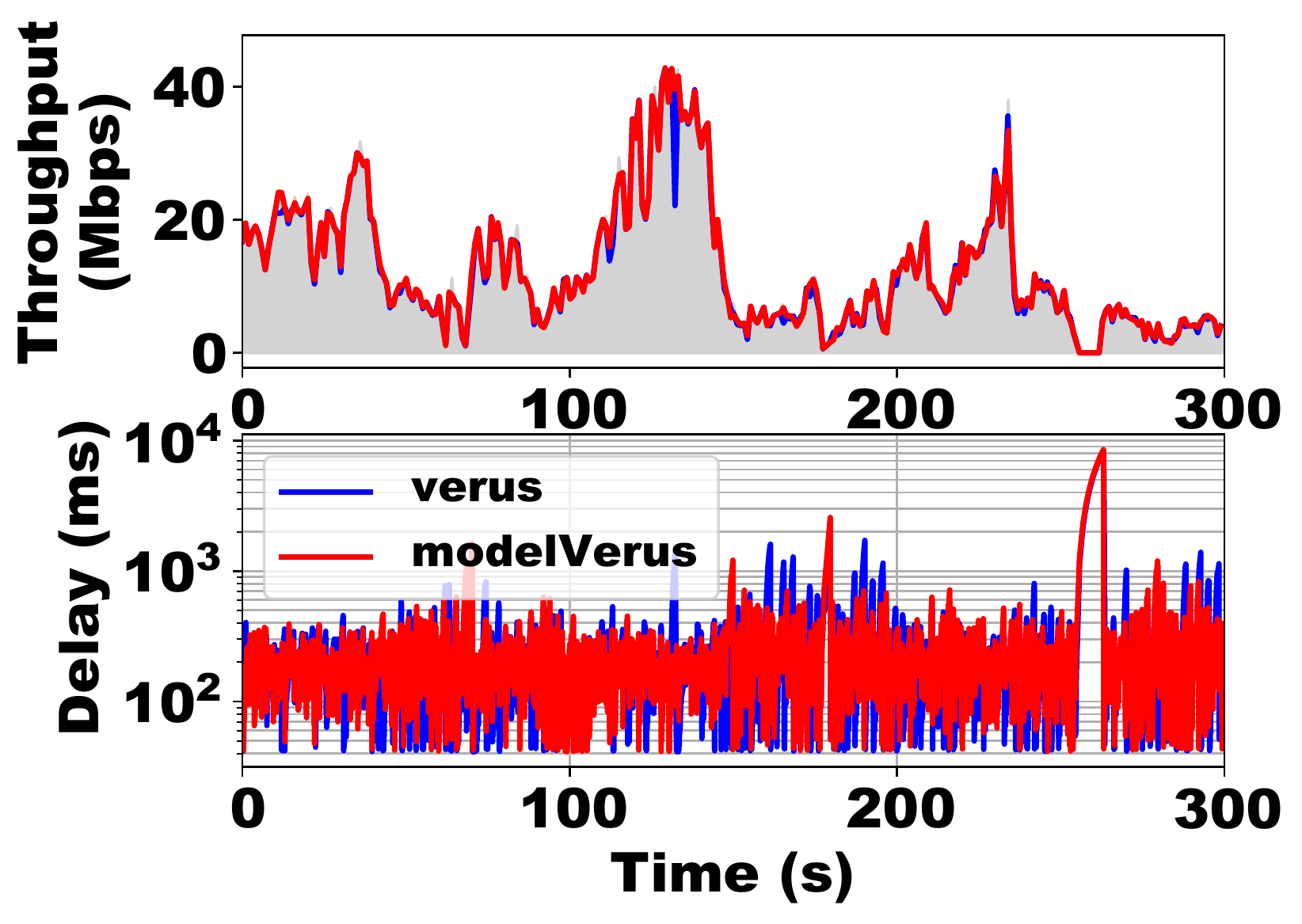}}
        \subfloat[\label{fig6:b}PDF]{\includegraphics[width=0.175\textwidth]{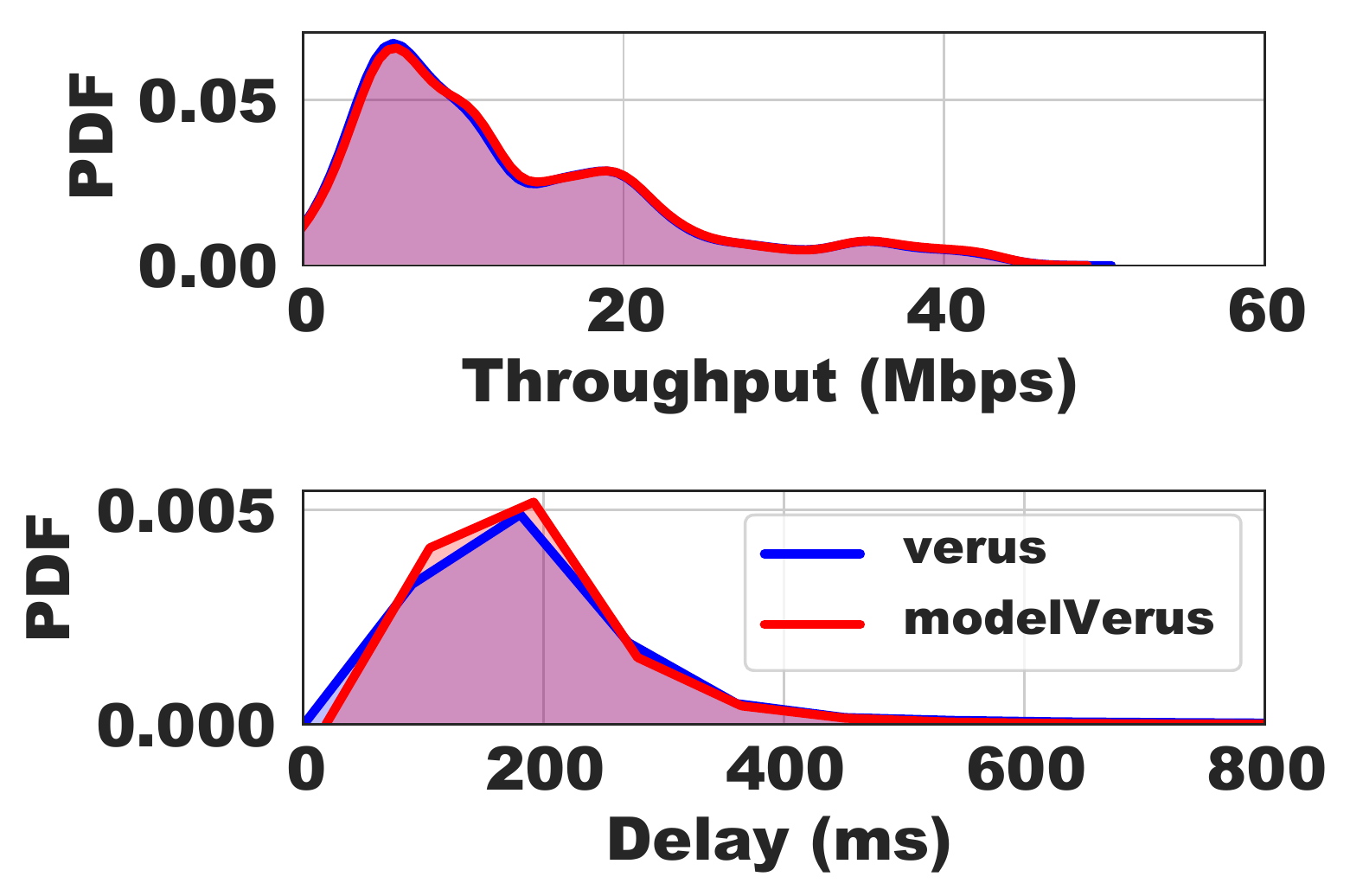}}
        \subfloat[\label{fig6:c}Population]{\includegraphics[width=0.155\textwidth]{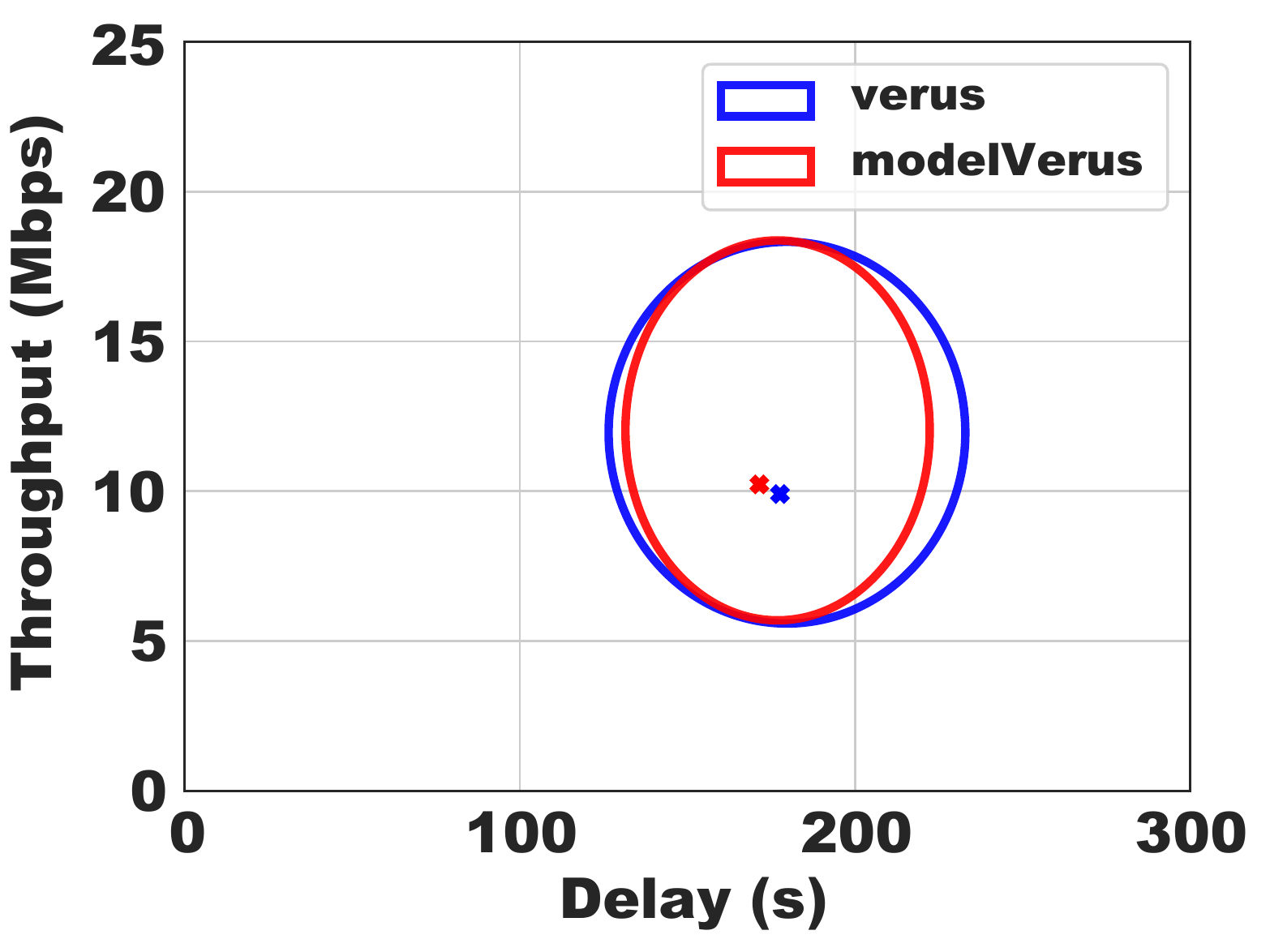}}
        \end{multicols}
        \caption{Verus 4G Verizon}
        \begin{multicols}{3}
        \subfloat[\label{fig7:a}instantaneous]{\includegraphics[width=0.17\textwidth]{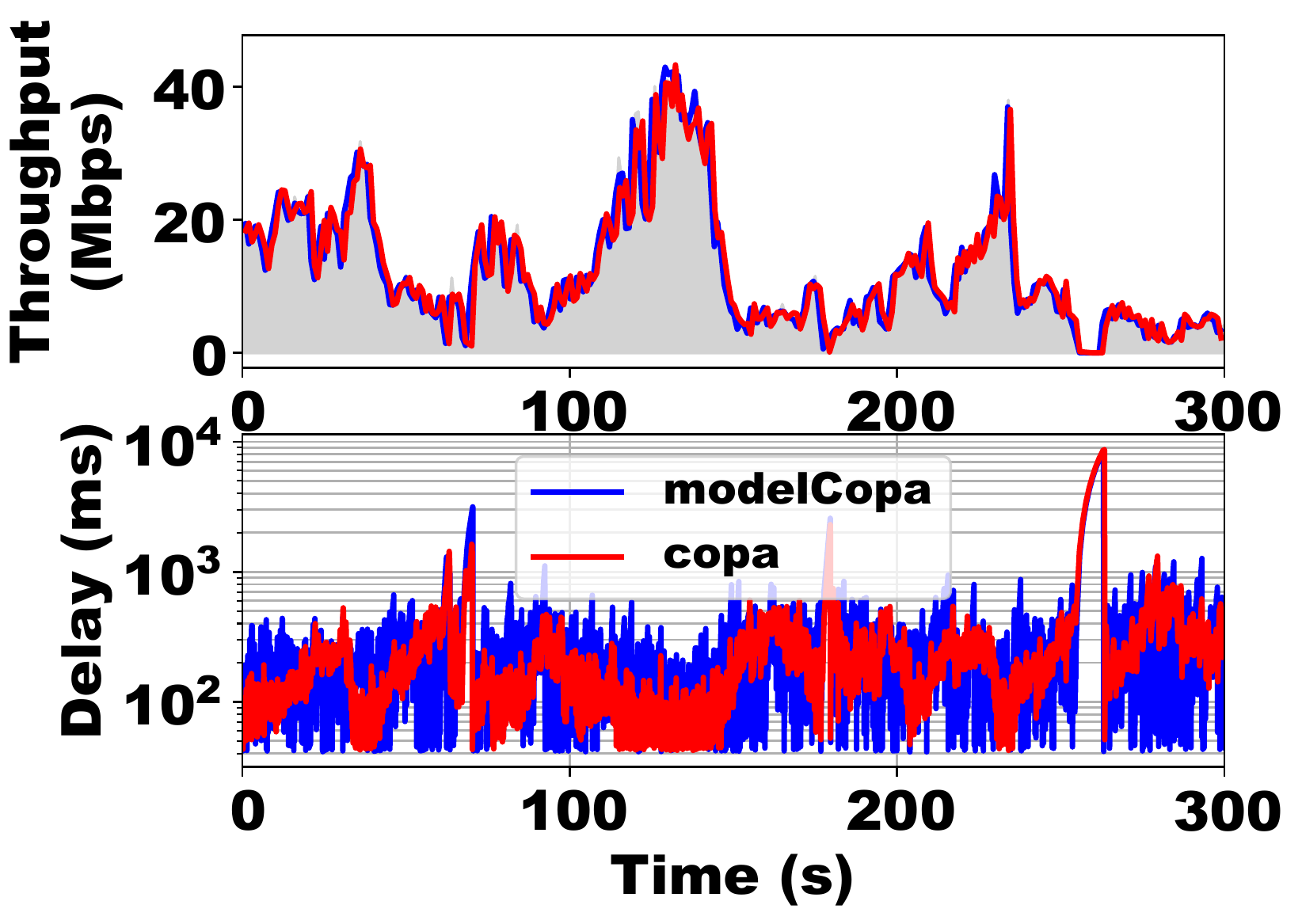}}
        \subfloat[\label{fig7:b}PDF]{\includegraphics[width=0.175\textwidth]{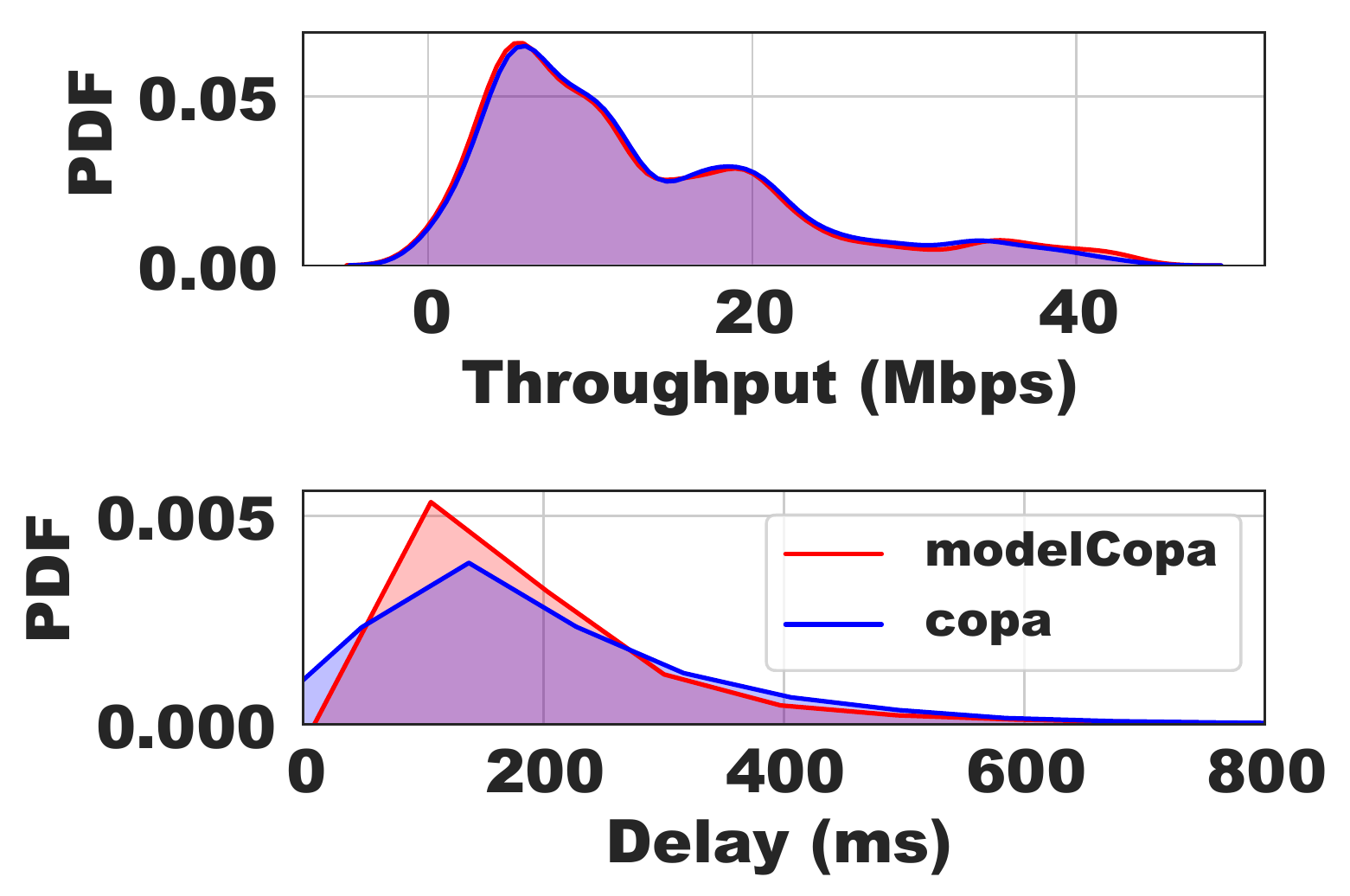}}
        \subfloat[\label{fig7:c}Population]{\includegraphics[width=0.155\textwidth]{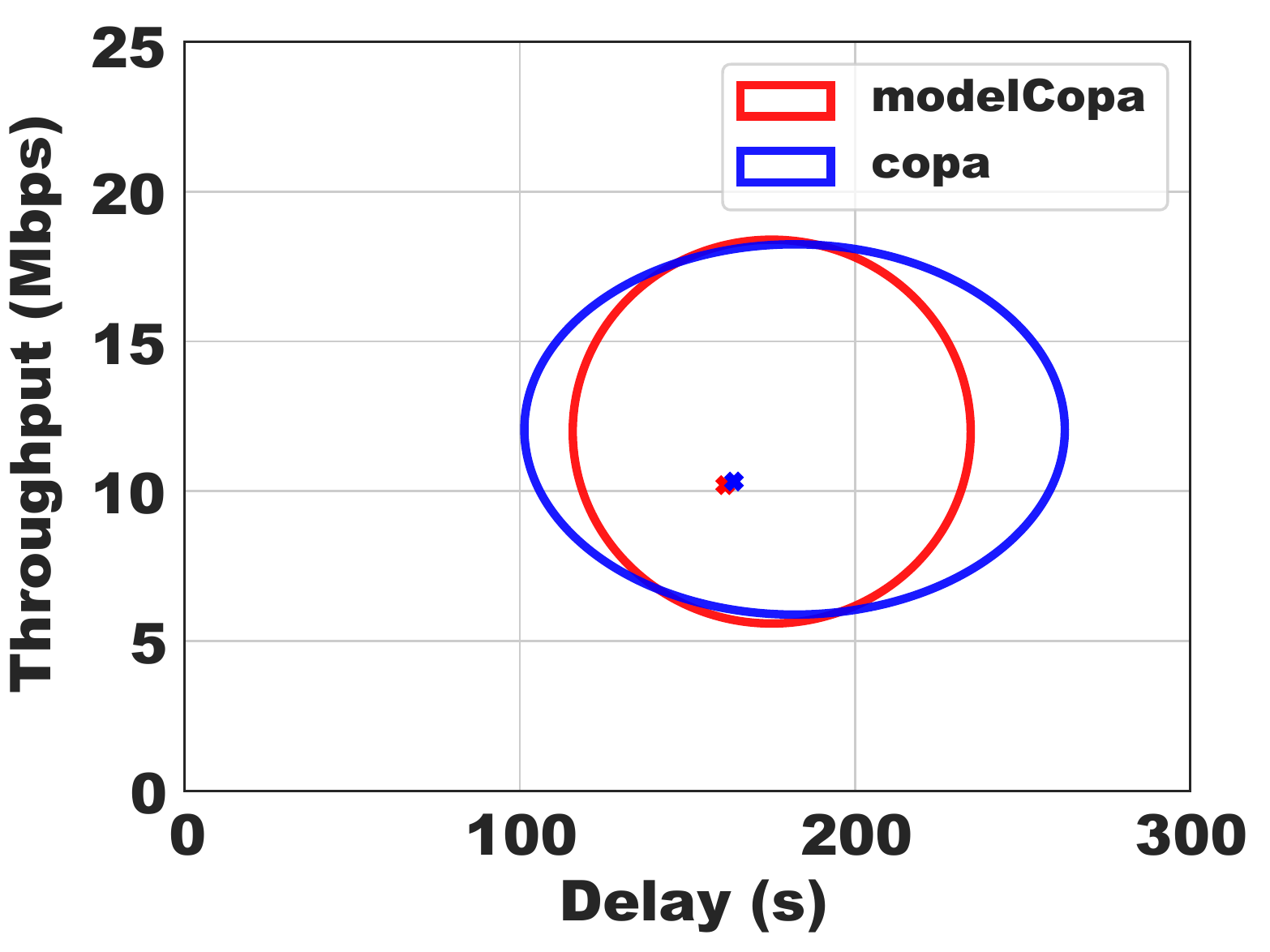}}
        \end{multicols}
        \caption{Copa 4G Verizon}
    \end{multicols}
\end{figure*}

\subsection{\name Implementation}
We implemented a generic sender and receiver in C that takes a transition matrix as an input and uses the matrix to decide the next sending window. The sender uses UDP as the transport protocol. It includes calculating the packet delays based on the incoming ACKs and uses the delay to determine the sending window size after each epoch. Epoch time is when the algorithm updates the congestion window. Algorithm~\ref{alg:gmcc_pseudo_code} outlines the \name control loop. The model algorithm identifies the next sending window $\hat{w}_{i+1}$ in every epoch, obtained from the transition matrix, where a row within a quadrant of the matrix represents all possible values for the future sending window. \name first identifies the operating quadrant through the row and column index. The row index is taken from the previous delay $\hat{d}_{i}$, and the column index from the current delay $\hat{d}_{i+1}$ (inferred from the incoming ACKs). Once the operating quadrant is identified, a particular row within the quadrant can be determined by the previous sending window $\hat{w}_{i}$. This row represents all possible sending windows decisions for the next epoch, each associated with a specific probability value. To decide the next sending window, \name draws a random number (between 0 and 1) to determines the closest matching sending window. This process is a guided random-walk within the state transition probability matrix. If the values are outside the matrix dimensions, the next sending window is determined by a multiplicative increase/decrease to the current window to force it back to the matrix bounds ($c_1$ and $c_2$).
\begin{algorithm}[h]
\caption{\name pseudo-code}
\label{alg:gmcc_pseudo_code}
\begin{algorithmic}[1]
\While {TRUE}
\State Compute $\hat{d}_{i+1}$ from ACKs
\If {$\hat{d}_{i+1} < \hat{d}_{min}$}
\State (Increase $\hat{w}_{i+1}$ using const. multiplier $c_1>1$)
\State $\hat{w}_{i+1} \gets \hat{w}_{i} * c_1$ 
\ElsIf {$\hat{d}_{i+1} > \hat{d}_{max}$}
\State (Decrease $\hat{w}_{i+1}$ using const. multiplier $c_2<1$)
\State $\hat{w}_{i+1} \gets \hat{w}_{i} * c_2$ 
\Else
\State $\text{Determine matrix quadrant} \gets \hat{d}_{i}$ and $\hat{d}_{i+1}$
\State $\text{Determine row within quadrant} \gets \hat{w}_{i}$
\State $\hat{w}_{i+1} \gets \text{Randomly choose next state}$ using transition probabilities in the chosen row
\EndIf
\State sleep(epoch) \Comment {epoch depends on the algorithm}
\EndWhile
\end{algorithmic}
\end{algorithm}
\section{Evaluation}
\label{sec:eval}
We evaluated two CC protocols as a proof-of-concept of \name: Verus, and Copa. These protocols are modeled through the training phase by generating the model transition matrix. The training is done using a set of 1000 different cellular traces (collected from real-world 3G/4G networks) that cover a wide range of network scenarios. To replay these traces, we used the MahiMahi~\cite{netravali2015mahimahi} linkshell network emulator. We used a different set of cellular traces for testing, taken from several previously published papers:
\begin{itemize}[noitemsep,topsep=0pt,parsep=0pt,partopsep=0pt]
  \item 4G Verizon: taken from~\cite{winstein2013stochastic} and represents a recorded channel over Verizon's 4G network in the US.
  \item Highway: taken from~\cite{zaki2015adaptive}, it represents a channel over a 3G network in the UAE while driving on a highway.
  \item Rapidly changing network: inspired by~\cite{dong2015pcc}, this trace represents a network with a highly fluctuating channel, where the capacity varied randomly every 5~seconds.
\end{itemize}

We wanted to evaluate how well a model representation of an algorithm can track the throughput and delay of the native algorithms when run on the same network traces. This section shows the results for the \name versions of Verus and Copa. 
For each protocol, we demonstrate the temporal variations of the original protocol against the \name version of the protocol for a snippet of a $300$ second run in one of the three scenarios in Figure~\ref{fig2:a}, \ref{fig3:a}, \ref{fig4:a}, \ref{fig5:a}, \ref{fig6:a}, and \ref{fig7:a}. The results show that across both protocols (Verus and Copa), the \name models (represented in red) are able to accurately track the throughput of the native protocol (represented in blue) temporally. In addition, the \name models are able to temporally track the delay behavior of these protocols. To quantify that the \name models statistically matches the characteristics of the original protocols, we computed the Probability Density Function (PDF) of the throughput and delay for both Verus and Copa respectively. Figure~\ref{fig2:b}, \ref{fig3:b}, \ref{fig4:b}, \ref{fig5:b}, \ref{fig6:b}, and \ref{fig7:b} shows the PDF comparisons. It can be seen that the \name throughput distributions match the native ones perfectly.

In summary, we observe that \name have the ability to accurately track the throughput behavior of the two used protocols across highly variable network conditions, evident by the results of Figure~\ref{fig2:c}, \ref{fig3:c}, \ref{fig4:c}, \ref{fig5:c}, \ref{fig6:c}, and \ref{fig7:c}. The figures show the overall summary comparing different values of the results population. Each of the \name model and the native protocol is depicted by a circular shape representing the operational region of the protocol circumscribed by the 25\% and 75\% percentile of the obtained throughput and delay, where the crosses (x) indicate the median values. The lower and upper part of the shape represents 25\% and 75\% of the throughput, respectively (y-axis), whereas the left and right part of the shape represents the 25\% and 75\% of the delay, respectively (x-axis). Results show that \name is capable of achieving quite similar statistical performance in terms of delay and throughput with a slight delay penalty not exceeding 5\% (i.e., in the rapidly changing channel).
\section{Related Work}
\noindent\textbf{CC for cellular networks:} conventional loss-based TCP variants, in particular Cubic~\cite{ha2008cubic}, performs poorly in cellular networks. This is due to the high sending rate that fills up the buffers causing a bufferbloat~\cite{gettys2011bufferbloat}. Bufferbloat is detrimental to the performance of delay-sensitive applications like video calling. This has led to newer delay-based CC protocols like Sprout~\cite{winstein2013stochastic}, and Verus ~\cite{zaki2015adaptive} that are specifically designed in the context of cellular channels. Sprout focuses on reducing self-inflicted queuing delays, and Verus creates a balance between the packet delays and the throughput. Recently, PCC Vivace~\cite{dong2018pcc}, which followed PCC~\cite{dong2015pcc}, has shown to react well to changing networks while alleviating the bufferbloat. PCC Vivace leverages ideas from online (convex) optimization in machine learning to do rate control. LEDBAT~\cite{RFC6817} is another delay-based CC algorithm developed for BitTorrent and other bulk-transfer applications that had limited adoption. 
BBR~\cite{cardwell2016bbr} was recently proposed by Google and has shown promising results over cellular networks. BBR uses the bottleneck link's round trip propagation and bandwidth to find CC's optimum operating point. 

\noindent\textbf{Applying machine learning to CC:} new CC protocols being proposed have complex control loops, which makes them harder to understand in the context of different network conditions. The recent development of CC protocols that employ machine learning (e.g., Remy~\cite{remy}, Vivace~\cite{dong2018pcc} and Indigo~\cite{yan2018pantheon}) have only compounded this issue (e.g., some of Remy's CC protocols employ rule tables with more than 100 rules). Weinstein et.\ al.\ (Remy)~\cite{remy}, Sivaraman et.\ al.~\cite{sivaraman2014} and P{\"o}tsch~\cite{potsch2016future} have provided different methodologies to model non-linear CC from a theoretical perspective. 

\noindent\textbf{Analyzing TCP behavior:} TCP and its variants have been thoroughly studied using the modeling, and analytical techniques ~\cite{olsen2003stochastic,samios2003modeling,wierman2003unified,padhye1998modeling,cardwell2000modeling}. A recent work called ACT~\cite{sun2019model} uses the concept of a guided random walk in the state space of implementation variables to find regions where the algorithm should never go, thereby indicating the existence of a possible bug in the implementation. Others also follow this approach of an automated model-guided method as well~\cite{jero2018automated} to explore the variable space in the implementation of a CC algorithm. Our modeling approach also uses a random walk, but our state space is limited to a delay and window variable, and our goal is not to reach unreachable points but to guide the model to follow the native algorithm it is modeling. 

\section{Discussion: Why MDI?}

\subsection{Visualizing Protocols}\label{sec:usecases:viz}
\begin{figure}[hbt]
    \centering
    \subfloat[\label{Verus-matrix}Verus]{{\includegraphics[width=0.238\textwidth]{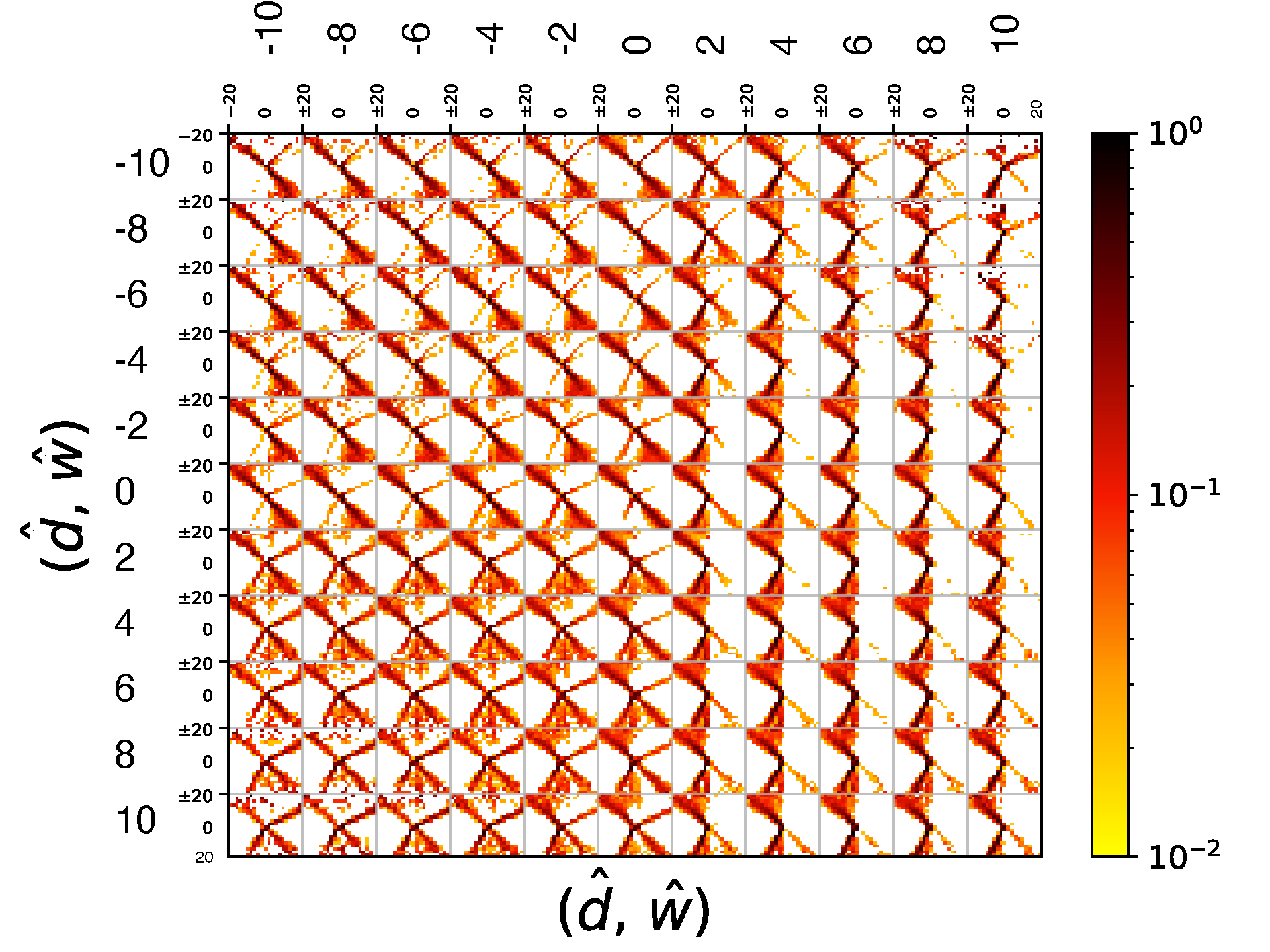}}}\hspace{0em}%
    \subfloat[\label{copa-matrix}Copa]{{\includegraphics[width=0.238\textwidth]{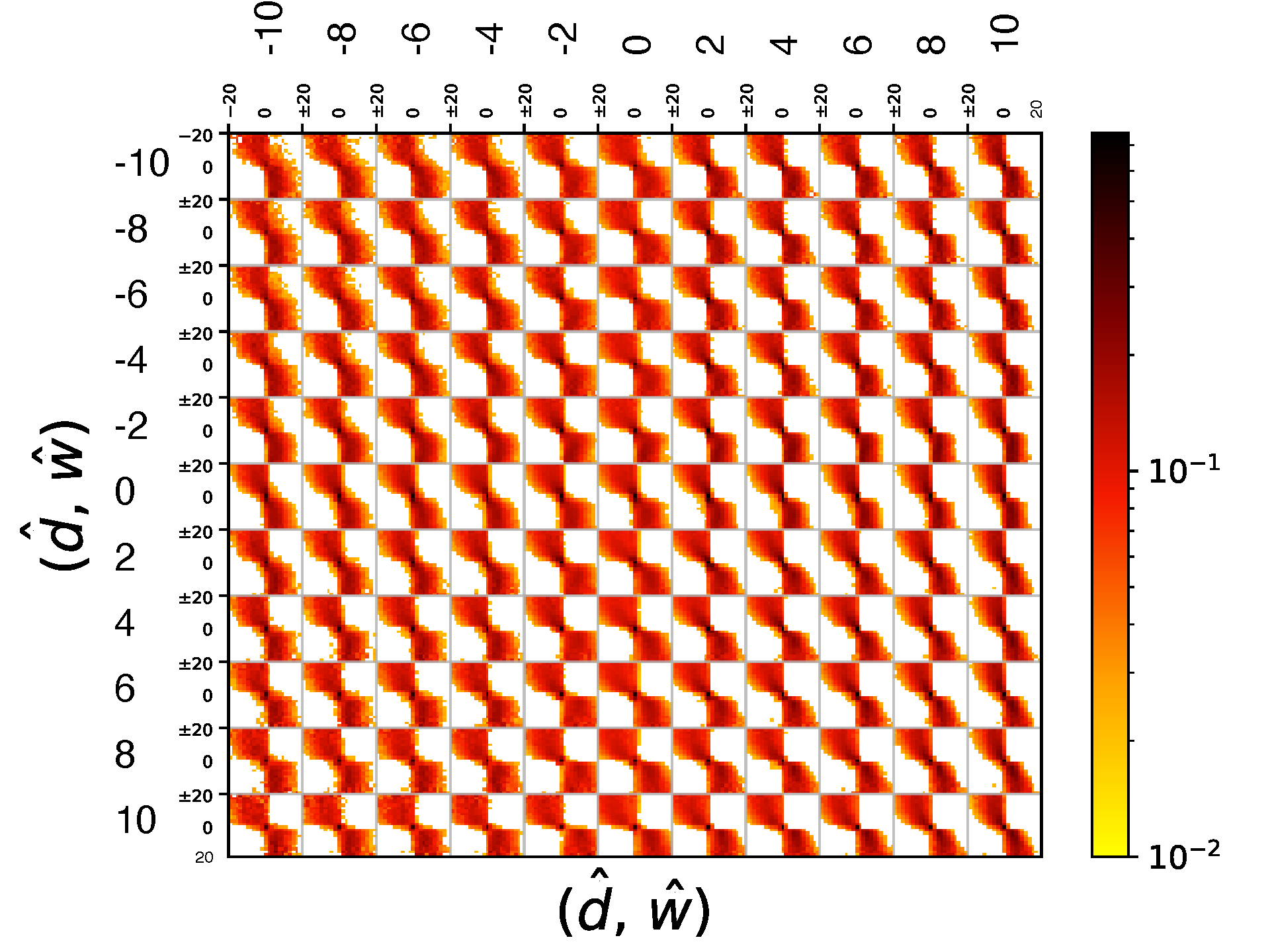}}}\hspace{0em}%
    \caption{\name transition probability matrices}%
    \label{mdi-matrix}
\end{figure}

The MDI transition matrix helps reason about the essence of the CC protocol behavior. These matrices represent the probability distributions across the transition space; it highlights which states the protocol mostly operates in. It also shows how the protocol is likely to behave under specific network changes, such as increased or decreased network delay. 
Verus and Copa's transition matrices (Figure~\ref{Verus-matrix} and~\ref{copa-matrix}) clearly show that the Verus matrix is less dense than Copa's, which means that Verus takes more decisive actions compared to Copa that tend to explore more. Each protocol shows a particular pattern reflecting the protocol's behavior; we call this the \textit{protocol fingerprint}. The sectors in the matrix represent different transitions for a specific change in packet delay. The relative delay and window ranges are determined from the training phase (the 1\% and 99\% of the observed increase/decrease population). 

The protocols' fingerprints reveal different characteristics of the protocol and how it reacts to various network changes. For example, the Verus transition matrix generally shows two distinct recurring patterns in the sectors: one on the left side of the matrix and the other on the right side. We can see that the right side pattern mainly contains window decrease probabilities. This is consistent with Verus's design, where if the observed delay increases, Verus lowers the sending rate by moving the operation point down the delay profile curve.
\begin{figure*}[th!]

    \begin{multicols}{2}
        \centering
        \subfloat[][Verus]{\includegraphics[width=0.5\textwidth]{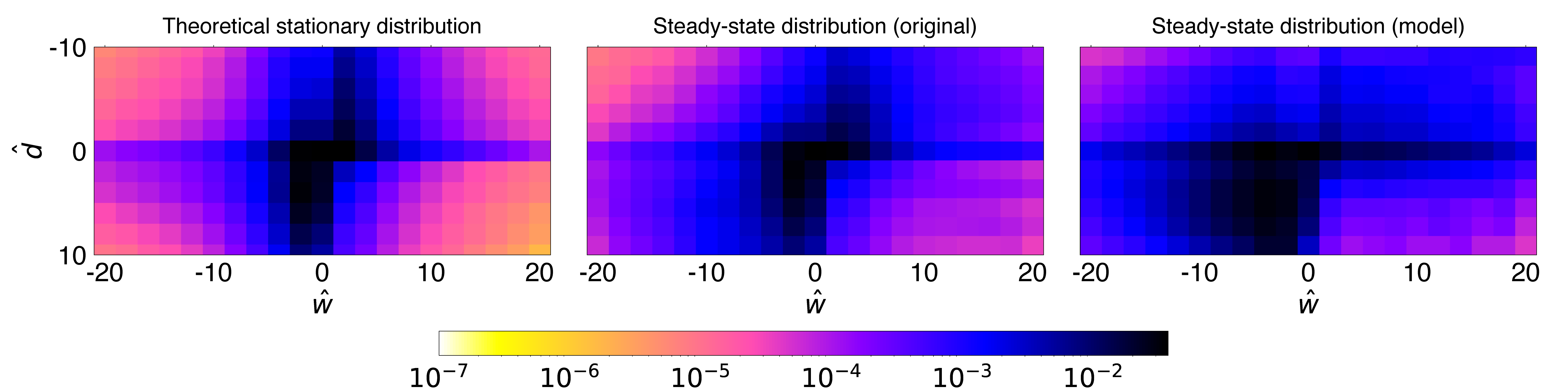}}
        \subfloat[][Copa]{\includegraphics[width=0.5\textwidth]{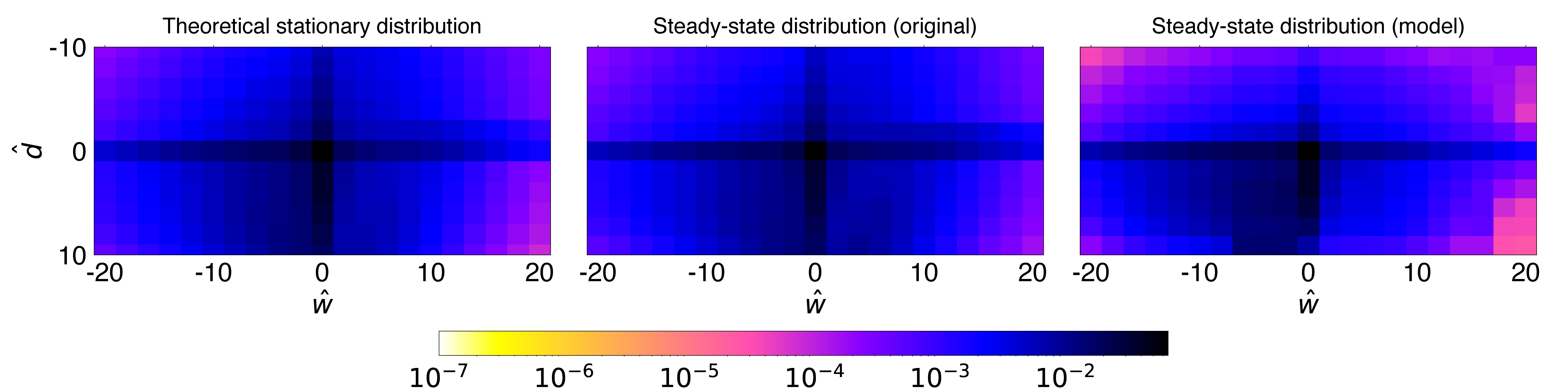}}
    \end{multicols}
    \caption{Comparison between the theoretical stationary (probability) distribution of the Markov chain model (left) that is trained on the training set of traces vs. the empirical distribution over the state space after mixing time for both the original and model versions of both protocols on the real-world test traces. These are for mixing time threshold ($\epsilon$) $10^{-3}$.}
    \label{fig:statdistrs-heatmap}
\end{figure*}


However, the left side pattern consists mainly of a diagonal line from the upper left corner down to the lower right corner. Additionally, the pattern also has an anti-diagonal, which becomes more dominant, moving down the sectors (i.e., when the delay feedback increases). This gives another insight to Verus. If a decrease in the previous delay is observed, it tends to continue alongside the same previous decision, extending the last window to decrease or increase. However, suppose Verus finds a delay-decrease with a prior increase in the delay. There is a higher probability that it might increase the window in the next decision despite the window decrease in the previous epoch. This confirms Verus's exploration behavior, where, in case of a delay reduction, it tends to increase the window to explore the channel variations immediately.

On the other hand, Copa's transition matrix shows that the matrix's right side sectors show almost the same pattern, with substantial probabilities in the upper left and lower right corners of the sectors and nearly no values in the top right or lower left edges. This means that regardless of the previous delay values or the severity of the observed delay values' increase, Copa tends to repeat its last epoch decision. For example, if Copa reduces the window, it will continue doing so in the next epochs. This is unlike Verus, where it tends to minimize the window in case of an observed delay increase. Looking at Copa's matrix's left sectors, we see that it has a similar pattern to the right side sectors with additional values in the upper right corner. These values become less dominant when moving down from the top to the bottom sectors. The sector's upper right corner represents increasing the window despite a reduction in the previous epoch. Like Verus, Copa tends to increase the window by observing a delay reduction, and the severity of exploring increases when the previously observed delays are decreasing.

\subsection{Convergence}

Using our Markov formulation, we can provide convergence guarantees as strong as the original protocols, using properties of convergence of Markov chains. Before presenting our results, we briefly review some necessary notations and definitions regarding Markov chains and convergence.

\textbf{Markov chains and Mixing times:} Every Markov chain can be represented as a transition matrix $P$, where the entry $p_{ij}$ represents the probability of transitioning to state $j$ from state $i$. Suppose $\mu^{(t)}$ is row vector that represents a probability distribution over the state space at a time $t$. Then at $t+1$, the distribution over the state space is given by $\mu^{(t+1)} = \mu^{(t)} P$. If the initial distribution at $t = 0$ is given by $\mu^{(0)}$, then we have from above that $\mu^{(t)} = \mu^{(0)} P^t$. The limiting distribution $\lambda$ is the limit of $\mu^{(t)}$ as $t \to \infty$. If a unique limiting distribution exists, then it equals the \emph{stationary  distribution}, which is the row vector $\pi$, such that $\pi P = \pi$. It is computed as the left eigenvector of the transition matrix corresponding to the largest eigenvalue~\cite{Norris1997}. The \textit{mixing time} of a Markov chain, $t_{\mathrm{mix}}$, is the time $t$ to convergence from an initial distribution $\mu^{(0)}$, i.e., when the probability distribution $\mu^{(t)}$ over the state space is sufficiently ``close'' to the stationary distribution $\pi$ that they are indistinguishable from one another. Any random walk process in a finite Markov space is associated with a finite mixing time~\cite{aldous1983random}. To obtain a conservative estimate, we define mixing time as the maximum convergence time starting from all possible initial states.

\textbf{Observations:} In our context, the state space comprises of the Cartesian product of 11 states in the delay space $\langle\hat{d}\rangle$ and 21 states in the window space $\langle\hat{w}\rangle$, a total of 231 $(\hat{d},\hat{w})$ tuples. If the start state is $i$, then the initial distribution $\mu^{(0)}$ is a one-hot vector, with 1 at the location corresponding to state $i$ and 0 everywhere else. Then, at every iteration $t$ (equivalent to an RTT), we compute $\mu^{(t+1)} = \mu^{(t)}P$, and declare convergence at time $t_{\mathrm{mix}}$ when the maximum element-wise difference between $\mu^{(t_{\mathrm{mix}})}$ and $\mu^{(t_{\mathrm{mix}} + 1)}$ is less than a certain defined threshold ($\epsilon$). We compute mixing times for three different thresholds: $10^{-3}$, $10^{-5}$ and $10^{-7}$. The last is chosen as it approximately equals the machine epsilon for 32-bit float. Table~\ref{tab:mixingtimes} shows the mixing times (in RTTs) obtained from the transition matrix for both protocols.

The heatmaps in Figure~\ref{fig:statdistrs-heatmap} show the theoretical stationary distribution computed using the Markov chain transition matrix trained over a training sample of 1000 traces, compared with the empirical distribution of states \emph{after convergence} (i.e., the mixing time) of the original protocols and the model versions over a separate testing sample of 60 cellular traces. The heatmaps are displayed over the two-dimensional $(\hat{d}, \hat{w})$ state space. The fact that these distributions match very closely is a robust result that our Markov model versions of the protocols are very close approximations of the original protocols. Table~\ref{tab:convergence} shows the closeness of the two distributions in terms of the Kullback-Leibler Divergence~\cite{Kullback1951} of the two distributions. The KL Divergence is a measure of how well one distribution approximates another. The closer the KL Divergence is to zero, the better the approximation. The table also additionally shows a simple maximum element-wise absolute difference between the two distributions. From the heatmap plots and these numbers, we observe that the model allows us to analyze the original protocols' convergence properties, which are a challenging proposition for delay-based protocols due to complex non-linear control loops.
\begin{table}
  \centering
  \scalebox{0.8}{%
  \begin{tabular}{lccc}
    Protocol & $\epsilon = 10^{-3}$ & $\epsilon = 10^{-5}$ & 
    $\epsilon = 10^{-7}$ \\
    \midrule 
    Verus & 24 & 55 & 85 \\
    Copa & 8 & 24 & 41 \\
  \end{tabular}
  }
  \caption{Mixing times (in RTTs) for both protocols, calculated from the Markov model.}
  \label{tab:mixingtimes}
\end{table}

\begin{table}
  \centering
  \scalebox{0.8}{%
  \begin{tabular}{lcc}
    Testing protocol & $D_{KL}(P||Q)$ & $\max|P-Q|$ \\
    \midrule 
    Copa & 0.017 & 0.004 \\
    Model Copa & 0.147 & 0.01 \\
    Verus & 0.101 & 0.02 \\
    Model Verus & 0.773 & 0.054 \\
  \end{tabular}
  }
 \caption{KL Divergence of the steady-state distribution of the states ($Q$) in the testing set after mixing time w.r.t. the stationary distribution ($P$) computed from the Markov model.}
  \label{tab:convergence}
\end{table}

\section{Conclusions}

This paper describes the MDI framework that can approximate delay-based protocols' behavior and potentially help visualize protocol behavior, understand convergence properties, and derive a model-based protocol replacement. 
We hope that this Markov modeling approach provides a new lens for understanding delay-based congestion control algorithms' behavior on highly variable networks.  
In future work, we hope to extend this framework to understand the behavior of a broader array of protocols, analyze fairness properties of MDI protocols and explore alternative state-space protocol representations within MDI.

\begin{acks}
	The work done by the authors Talal Ahmad, Shiva Iyer and Lakshminarayanan Subramanian in this paper was supported by a  \grantsponsor{HR001117C0048}{Defense Advanced Research Projects Agency (DARPA)}{https://www.afcea.org/content/Blog-darpa-awards-three-dispersed-computing-contracts} contract \grantnum{HR001117C0048}{HR001117C0048}. Any opinions, findings and conclusions or recommendations expressed in this material are those of the author(s) and do not necessarily reflect the views of DARPA.
\end{acks}

{ \balance
{


    \bibliographystyle{ACM-Reference-Format} 
    \bibliography{references}
}
}

\end{document}